\documentclass[aps,pra,twocolumn,groupedaddress]{revtex4}
\usepackage{graphicx}
\usepackage{amsmath}
\usepackage{amssymb}

\def\gsim{\mathrel{\rlap{\lower4pt\hbox{\hskip1pt$\sim$}}\raise1pt\hbox{$>$}}}

\begin{document}

\title{Pseudogaps in strongly interacting Fermi gases}

\author{Erich J. Mueller}
\affiliation{Laboratory for Atomic and Solid State Physics, Cornell
University, Ithaca NY 14853}

\date{\today}

\begin{abstract}
A central challenge in modern condensed matter physics is developing the tools for understanding nontrivial yet unordered states of matter.  One important idea to emerge in this context is that of a ``pseudogap": the fact that under appropriate circumstances the normal state displays a suppression of the single particle spectral density near the Fermi level, reminiscent of the gaps seen in ordered states of matter.  While these concepts arose in a solid state context, it is now being explored in cold gases.  This article reviews the current experimental and theoretical understanding of the normal state of strongly interacting Fermi gases, with particular focus on the 
phenomonology which is traditionally associated with the pseudogap.
%evidence for/against a pseudogap.
\end{abstract}

\maketitle

\section{Introduction}

In this review I describe attempts to understand the %single particle spectra 
normal state properties
of strongly interacting Fermi gases (typically $^{40}$K or $^{7}$Li) cooled to quantum degeneracy, with particular
focus on the single particle spectrum.
At sufficiently low temperatures, $T<T_c$, the atoms pair up to form a superfluid.  Given that it takes energy to break the pairs, there is no way to remove a particle without adding energy to the system: The single particle spectrum contains a gap.  This gap is extremely important, and almost all properties of a superconductor/superfluid either follow from this gap or from the behavior of the order parameter.  A key question is what happens to this gap when $T>T_c$.   In high temperature superconductors,  vestiges of the gap remain in the normal state.  
%Although there are serious doubts about the connection of this spectral feature to superconductivity, 
These observations caused researchers to question if such ``pseudogaps" are a generic feature of strongly interacting fermions. The path to answering this question brings us through one of the most important areas of modern condensed matter physics research: understanding correlations in disordered states of matter.  As I will explain, experimental studies of ultracold Fermi gases have played an important role in solidifying our understanding of these questions.
%Related questions appear in studies of materials such as **insert material**, a frustrated spin system,  and YBCO, a high temperature superconductor.  In particular, in the high temperature superconductors, vestiges of the gap remain in the normal state.

In section~\ref{challenge} through \ref{crossover}, I develop the themes which motivate these studies. 
 In section \ref{experiment} and \ref{theory} I review the experimental and theoretical work.  In \ref{summary} I summarize the results, and outline the prospects for the future.

There are a number of books and reviews on the subject of strongly interacting Fermi gases \cite{chevymora,feshbachreview,compilations1,compilations2,compilations3,bloch,giorginireview,BCSBECreview,tormareview,chen,levinxx,levinhulet,melo,ketterlereview,jinreview}, each with their own focus.  Of particular note is Chen and Wang's recent review of pseudogap physics in Fermi gases with an emphasis on a theoretical approach which they helped develop
 \cite{chen}.    Chen was also an author of an earlier review with a similar focus \cite{levinxx}.  While there inevitably will be some overlap, my perspective will be broader, and hence complementary.  
 
In order to keep the discussion focused, I will restrict the discussion to three-dimensional systems.  Analogous physics also is found in quasi-two dimensional gases \cite{twod}.  I will also avoid the very interesting subject of ``imbalanced" gases, where there is a finite spin polarization \cite{imbalanced,chevyreview}

%

% the story is  more complicated. 

%
%Cold gases are allowing us to controllably study the properties of collections of many interacting particles.  A particularly important direction for this research is studying how correlations manifest themselves in disordered states of matter.
%%One of the many interesting directions of this research

%Here I will introduce the major theme: understanding correlations in normal states and how they show up in spectra.  A concrete example is the behavior of excitations in the BCS-%BEC crossover of dilute Fermi gases:  In the BEC regime the normal state excitations are all bosonic.  In the BCS regime they are fermionic.  How does one smoothly join these %behaviors?  %Another important paradigm is para-conductivity, the observation that the conductance of a normal metal is enhanced near a superconducting transition.

% Can we say something about Tan?

\subsection{The challenges of understanding normal states}\label{challenge}

It is often much more challenging to understand ``disordered" states of matter than ``ordered" states.  A good example comes from comparing a classical liquid and a classical crystal.  The simplest cartoon of a crystal involves pinning down the exact location of each particle.  This is a reasonable starting point, and is easily expanded on to make predictions (for example, one can connect the particles with ``springs" to model the low energy excitations).  The simplest cartoon of a liquid is that each particle is equally likely to be at any place in space.  While a reasonable model for a gas, this does not give a good starting point for a liquid, where the individual molecules are in constant contact.  The key to a liquid is that the atomic positions are {\em correlated}, but that they are not {\em ordered}.  This tension between having structure, but not too much structure, makes the resulting theory more complex.

%To describe the ground state of the crystal, one need only specify the location of the particles in a unit cell, and give the vectors which generate the lattice.  The low energy excitations can be calculated from a simple model of springs connecting those particles.  Bulk properties, such as elastic moduli, are readily derived from these spring constants.  The liquid, however, is more complicated.  A typical description might involve specifying the "correlation function" $g(r)$, which gives the probability that any two particles are separated by the distance $r$. 

There are many areas of modern
condensed matter physics where intense effort is devoted to understanding disordered quantum mechanical state of matter.  These include: Frustrated spin systems, which sometimes have disordered ground states dubbed ``spin liquids" \cite{spinliquid}; high temperature superconductors, where the normal state shows a range of behaviors, described by terms such as ``pseudogap" and ``strange metal" \cite{randeria,levin,timusk,lee,leereview};  multiferroics, where proximity to various ordered phases may lead to technologically useful properties \cite{multi}; quantum critical systems, where the proximity to a zero temperature phase transition drastically modifies the normal state \cite{sachdev}; and one-dimensional wires, where kinematic constraints and the high density of states at low energy suppress ordering \cite{giamarchi}.

What makes these normal states hard to model is the fact that they are not well described by noninteracting electrons (the analog of modeling a classical liquid by saying the positions of all the particles are at random locations).  One typically describes them as ``strongly correlated."  Because interactions are strong, perturbation theory about the noninteracting state fails.  One can use variational techniques \cite{jastrow, gutzwiller}, but optimizing these variational wavefunctions are far from trivial \cite{cyrus}, and even calculating the energies of these variational states is challenging.  Moreover, without insight into the physics, one has little hope of constructing an appropriate variational wavefunction.   Ab-initio methods, such as quantum Monte-Carlo \cite{mc}  hold promise, but often face technical difficulties.
%tend to break down as one approaches the most interesting regimes.  
%Other approaches, such as dynamical mean field theory \cite{dmft} are biased \cite{dmftfailings}.
These states are disordered, so there is no conventional  ``mean field theory" for describing them.

Given the difficulty of understanding the microscopics of these systems, it is important to develop phenomenological pictures of the emergent physics.  
Pseudogaps provide one natural organizing principle. 
%Some argue that the pseudogap plays a central role in the phenomenological models of the normal states of strongly interacting fermions.
%(defined precisely in sec.~\ref{pseudo}) has been identified as one of the key properties of the normal state of the cuprates.  What aspects of the pseudogap are unique to such transition metal oxides, and what aspects have more general importance?

%explFor example, when do you see 

%
%An important goal is to figure out what the emer

%For example, Other examples are spin liquids: disordered ground states of frustrated spin systems.

%Similarly, with a superconductor, the normal state can be more difficult to understan

%Despite this challenge, it is cruc

%In quantum mechanics describing a no

%%The quantum mechanics of many interacting particles can be extremely complicated.  

%In 19?? Landau wrote a key article which set the language which is used for thinking about states of matter and their properties.  He introduced the concept of an "order parameter," a quantity which is zero in the "disordered" state, and nonzero in the "ordered" state.  The classic example is the magnetization of a ferromagnet.

%
%Here I discuss why normal states pose more of a challenge than ordered states.
%%I'll bring up multiferroics to give a bit of motivation about how being near ordered states gives interesting and possibly useful properties.  I'll also mention quantum criticality and how phase transitions qualitatively impact the normal state.

\section{What is a pseudogap?}\label{what}

As a phenomenological concept, there is no sharp definition of a pseudogap, and one sees the term used in several different field.  In this section I give a brief description of superconducting order parameters, which then allows us to describe several proposed definitions.  

%\subsection{Superconductivity: Order parameters and spectra}\label{superconductivity}
%The principle context in which pseudogaps are discussed is superconductivity -- and it is from this field that the nomenclature arose.
The simplest model of a superconductor \cite{degennes} yields a single particle excitation spectrum
\begin{equation}\label{bcs}
E(k)=\sqrt{\left(\frac{k^2}{2m}-\mu\right)^2 +\Delta_s^2},
\end{equation}
where $k$ is the momentum of the excitation, $m$ the electron mass, $\mu$ the chemical potential, and the spectral gap
$\Delta_s$ is the energy of the lowest energy excitation. 
The order parameter of a superconductor is typically taken to be
%not only an important spectral feature, but it also quantifies the degree of pairing in the system, 
$\Delta_0=g \sum_k \langle a_{k\uparrow}^\dagger a_{-k\downarrow}^\dagger\rangle,$ where $g$ is a typical interaction energy (traditionally related to the electron-phonon coupling), and $a_{k\sigma}$ is the annihilation operator for an electron with momentum $k$ and spin $\sigma$.  This order parameter quantifies the degree of ``pairing" in the system, and it 
vanishes in the normal state.  The intuitive picture of the order parameter $\Delta_0$, is that it measures the fraction of fermions which are bound up in Bose condensed pairs.  
%Section~\ref{crossover} elaborates on this understanding.  
Within weak-coupling mean field theory, $\Delta_0=\Delta_s$, and the term ``gap" is used interchangably for each of these.

Generically $\Delta_s\neq \Delta_0$.  In fact, it is possible to have pairs ($\Delta_0\neq 0$) without a spectral gap ($\Delta_s=0$).  This occurs in strongly disordered superconductors \cite{woolf,ag}.  Conversely, there are many sources of spectral gaps which have no connection with superconductivity.

Thus %to someone new to the termonology, 
the phrase ``pseudogap" could naturally have two different meanings:  It could refer to $\Delta_s$, describing a spectrum which lacks a gap, but which has a suppression of the density of states near the Fermi surface.  Alternatively it could refer to $\Delta_0$, describing a system in which there is some precursor of pairing or in which there exists pairs which are not Bose condensed.  A common assumption is that these two meanings of ``pseudogap" are linked, and that the existence of superfluid precursors leads to a suppression of spectral weight.  Such precursors have been widely studied \cite{paraconductivity}.  One of the important features of the cold gas system is that it provides a setting in which one can investigate these links.

The nomenclature is even more muddied.  Some \cite{timusk} define a pseudogap to mean a that the spectrum is gapped in some regions of momentum space, but are ungapped in others.  Others \cite{levinxx} define a pseudogap by the condition that $\Delta_0=0$ but $\Delta_s\neq0$.  Some \cite{Tsuchiya} even define multiple ``pseudogaps" corresponding to multiple spectral features.  All of these definitions have merit, and most controversies about the ``existence" of pseudogaps \cite{chen} typically reduces to differences in definitions.  

The most widely used definition is that a pseudogap is a depression in the single-particle density of states near the Fermi energy.  Even with this definition, there is not a unique scale associated with the phenomenon.  Just taking one set of authors, Tajima et al. use $\tilde T^*$ to denote the highest temperature at which this depression appears \cite{ohashispin}.  They also introduce $T^*$ as the temperature at which the density of states at the Fermi surface is maximal.  Typically $T^*\gg \tilde T^*$.  This nomenclature is by no means standard.  For example, Magierski et al. use $T^*$ to denote the temperature at which a depression first appears in the density of states \cite{magierskigap}.

A second approach to defining a pseudogap involves fitting the excitation spectrum to Eq.~(\ref{bcs}).  The coefficient $\Delta_s$ then quantifies the pseudogap.  Within this approach one can have a depression in the density of states without having a pseudogap.   For example, take $\mu<0$ and $\Delta_s=0$.  The spectrum in Eq.~(\ref{bcs}) then has a gap of $-\mu$.   While it has the advantage of cleanly connecting to the ordered state, this fitting approach can seem overly rigid.  In particular, Eq~(\ref{bcs}) is often not a good model of the single particle spectrum, in which case this definition is meaningless.

To further understand this issue, one must note that in general  it is impossible to ascribe a one-to-one relation between momentum and energy: Interactions with other electrons, phonons, or impurities, mean that an electron will only have a momentum $k$ for a finite time.  Due to the energy-time uncertainty relationship this gives an uncertainty in the electron's energy.  Thus the natural way to describe the excitation energies is through the spectral density $A_k(\omega)$.  The probability $P$ that an excitation of momentum $k$ has energy between $\hbar \omega_1$ and $\hbar \omega_2$ is $P=\int_{\omega_1}^{\omega_2} A_k(\omega) d\omega/(2\pi)$.   In an ideal gas $A_k(\omega)= 2\pi \delta(\hbar\omega-k^2/2m)$ is non-zero only when $E=k^2/2m$.  Finite lifetime broadens this spectral function.   In the BCS theory, $A_k(\omega)$ has two branches, and is nonzero when $\hbar\omega=E(k)$ or $\hbar\omega=-E(k)$.  The former corresponds to excitation processes where one an unpaired particle, while the latter to one in which you a pair and a hole.

%two-branches are interpreted in terms of the two ways of adding a particle to the system:  One can either add a single unpaired particle, or one can add a pair, and a hole. 

\begin{figure}
\includegraphics[width=\columnwidth]{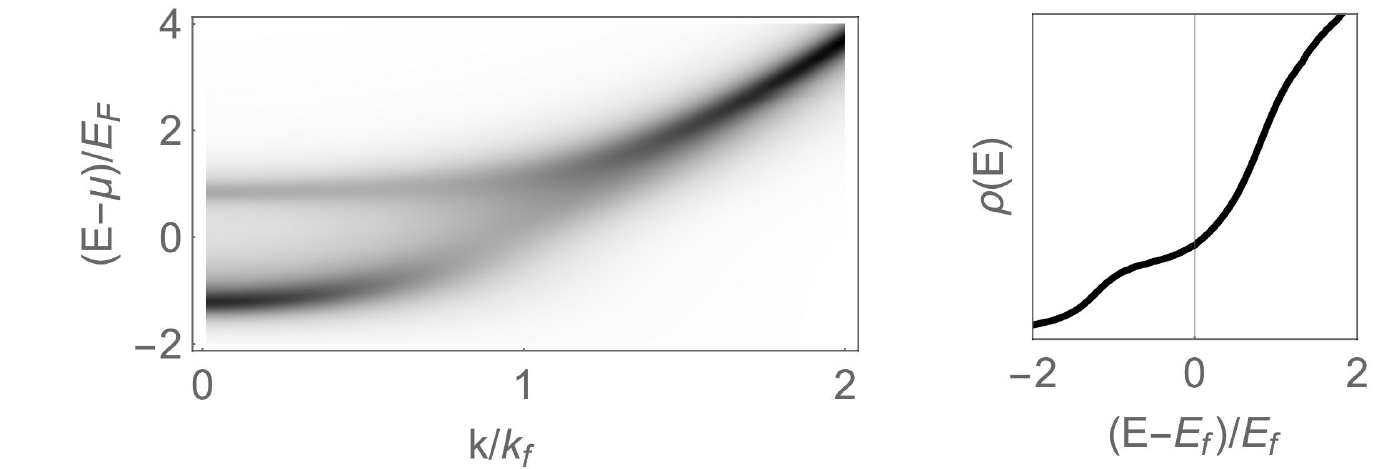}
\caption{Left: Spectral density $A_k(\omega)$ of a strongly interacting (unitary) Fermi gas at $T/T_c=2$, within the Nozieres and Schmidt-Rink T-matrix approaches described in sec.~\ref{diag}.  Dark colored regions correspond to higher spectral density, and indicate the relationship between energy $E=\hbar\omega$ and momentum $\hbar k$ in the single particle states.   Right: Corresponding density of states, showing a distinct dip near the Fermi level.} \label{pgfig}
\end{figure}

Figure~\ref{pgfig} shows a typical spectral density $A_k(\omega)$ in one model of the strongly interacting Fermi gas described in section~\ref{fg}.  Darker colors represent higher spectral density, and to the extent one can define a dispersion relationship, it should follow the darkest regions.  For this particular model (the Nozieres and Schmidt-Rink T-matrix approach, see sec.~\ref{diag}), and these parameters ($T/T_c=2$, $a_s=\infty$ -- see sec.~\ref{fg}), one has a distinct two-branch structure, but the dispersion is  not well described by Eq.~(\ref{bcs}).  Nonetheless, the density of states is suppressed near the Fermi level, and it would seem reasonable to say that this state has a pseudogap.  
Note, there are other approximations, such as the one developed by the University of Chicago group \cite{chien}, which yield spectral densities which are better fit by Eq.~(\ref{bcs}).  

To further add to the controversy, some advocate that one should reserve the phrase ``pseudogap" for the case where the pairing comes from many-body effects rather than two-body effects.  Others further stipulate that the temperature should be sufficiently low, or that the gas should be a ``non-Fermi liquid".  

This diversity of definitions is natural in a young field, and as time evolves consensus will develop.  
In this review I will take as expansive a definition as possible of the pseudogap, and use the term for both pairing and spectral features, without attaching any additional requirements.   This broad definition has  relevance to the widest range of systems.

\section{Where are Pseudogaps formed?}

While the main purpose of this review is to discuss the BCS-BEC crossover in cold Fermi gases, the importance of the subject is only clear by looking at some of the cannonical examples of pseudogaps.  In this section we discuss the cuprate superconductors and 1D charge-density-wave materials.

\subsection{The Pseudogap regime in the cuprate superconductors}\label{pseudo}
The concept of the pseudogap was developed in the normal state of ``underdoped" cuprate superconductors \cite{randeria,levin,timusk,lee,leereview}. The cuprate superconductors (such as La$_{2-x}$Sr$_x$CuO$_4$ and YBa$_2$Cu$_3$O$_{7-x}$) are doped antiferromagnetic insulators \cite{butch}.  The ``parent compound" formed when $x=0$ is magnetically ordered at low temperatures.  As one increases $x$ to a few percent, the magnetic transition temperature appears to drop to zero.   With increasing $x$, one finds  superconducting order.  the superconducting transition temperature  rises with $x$, peaking around $x=20\%$, denoted ``critical doping" $x_c$.  Any $x<x_c$ is refered to as ``underdoped."  At larger $x$ the superconducting transition temperature falls to zero.

In the underdoped cuprates, the pseudogap refers to a collection of phenomena which can be 
 interpreted as a reduction of the density of states for low energy excitations. 
For example, both spin susceptibility and spin relaxation are suppressed \cite{randeria}.  This suppression in the spin susceptibility/relaxation would, for example, be consistent with the electrons being bound up in singlet pairs -- ie. it is natural to interpret it as a precursor to superconductivity.  

The pseudogap has been seen in  spectral probes  (photoemission \cite{pe}, tunneling \cite{tn},  magnetic resonance \cite{nmr}), transport (optical and DC conductivity \cite{opticalcond,dccond}) and thermodynamic   probes (magnetic susceptibility \cite{susc}, specific heat \cite{sh}).  One of the more intriguing results was the observation that, in a magnetic field, applying a thermal gradient will induce a relatively large transverse voltage drop \cite{nernst,tannernst}.  This ``Nernst" effect has been interpreted as a sign of ``vortices" in the normal state, and hence a superconducting precursor.  

While it is natural to ascribe a connection between the various observations and superconductivity, the modern concensus\cite{hiddenarpes,hidden} seems to be that they are instead a signature of ``hidden order." Candidate orders include magnetic order, ``d-density waves", and ``electronic nematics."  Both scanning tunneling microscopy and angle resolved photoemission spectra reveal a ``two-gap" structure, where spectral suppression near the nodal direction are attributed to superconductivity and structures near the antinodal direction are attributed to the ``pseudogap".  The ``nodal gap" vanishes above $T_c$, while the ``antinodal gap" vanishes at a higher temperature, $T_{\rm pg}$.  Strong evidence of liquid crystaline smectic/nematic order has been seen in scanning tunneling microscopy \cite{lawler}.  

It is not clear if the hidden order competes with superconductivity, or enhances it \cite{hidden,friendfoe}.  Some argue that Superconductivity emerges from the quantum fluctuations found near the quantum critical point where the hidden order vanishes \cite{quantcrit}.  One field of thought is that the relationship between the various possible orderings is more complicated, and advocate the phrase ``intertwined order" \cite{intertwined}.  Interestingly, there are arguments that the phenomonology of the strongly interacting normal state is largely independent of the nature of the order \cite{nikolic}.

\subsection{Pseudogaps in 1D Peierls systems}\label{peierls}
It is useful to have a controlled model system to understand how spectral signatures of ordering can be found in an un-ordered state.  The simplest such example is the
%In a traditional weak-coupling BCS superconductor there is no pseudogap:  The normal state has no appreciable spectral signatures of superconducting order, and there are no significant pairing fluctuations.  A strong coupling superconductor is har
%
%A better understood  pseudogaps occurs in found in the 
1D electron gas: a system which is unstable towards forming a charge-density wave \cite{peierls}.  
A mean-field treatment of that instability parallels closely the BCS theory of superconductivity, and the single-particle spectrum is similar to that in Eq.~(\ref{bcs}).  In this mean-field theory, the electron density develops sinusoidal oscillations.  The electric field from this inhomogeneous charge density provides a potential, which reinforces the modulation.  The electron density is commensurate with the oscillations, and the electrons form a band insulator in this effective potential.  In this context, the spectral gap $\Delta_s$ appearing in Eq.~(\ref{bcs}) is associated with the band-gap of the self-consistent potential, and has no connection with superconductivity.

\begin{figure}
\includegraphics[width=\columnwidth]{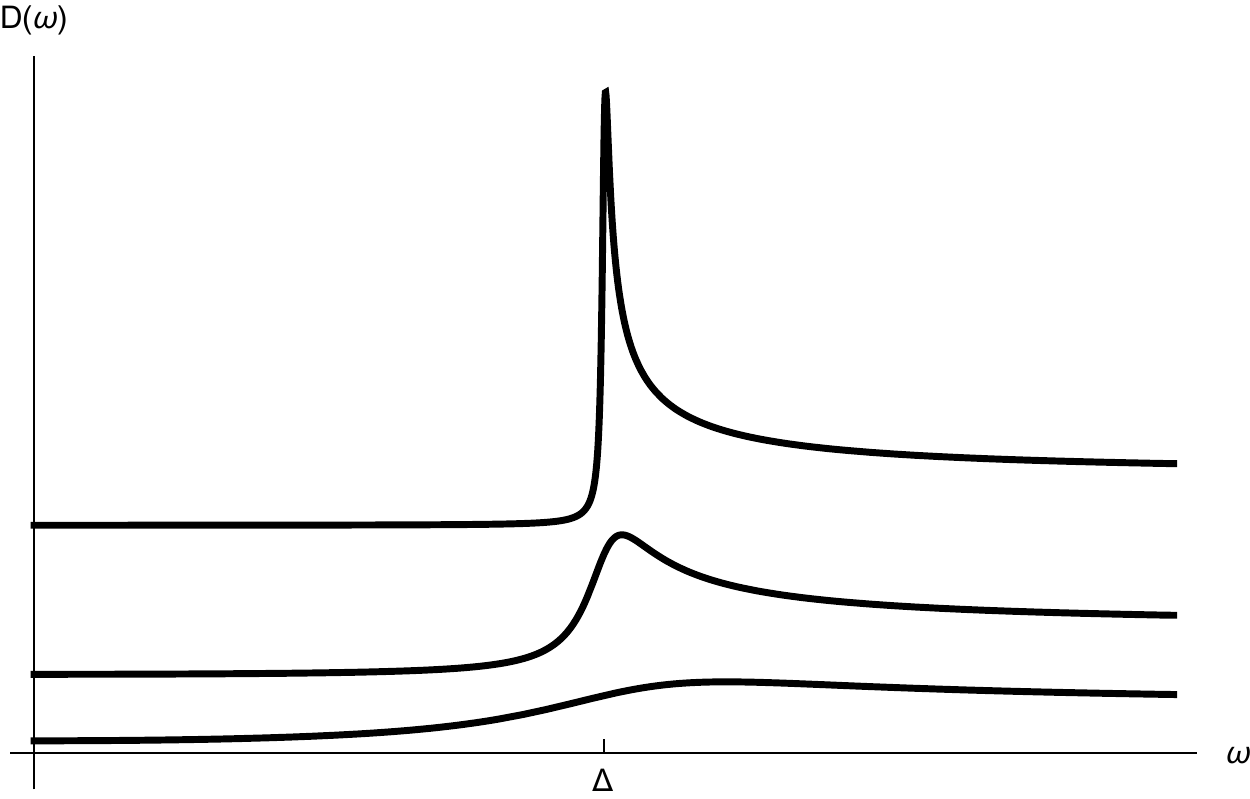}
\caption{Single particle density of states $D(\omega)$ of a 1D metal with a charge-density wave instability, as calculated by Lee, Rice, and Anderson \cite{lra}.  Fluctuations destroy any long-range order at finite temperature, removing the spectral gap, but leaving a pseudogap.  Bottom to top shows decreasing temperature. Curves are offset for clarity.} \label{leefig}
\end{figure}

While valuable, this mean-field theory is incomplete, and one expects no long-range order at finite temperature for this system \cite{mermin}.  As Lee, Rice, and Anderson \cite{lra} agued, and illustrated in Fig.~\ref{leefig}, when one includes fluctuations the spectral density no longer contains a gap, but there is a notable depression in the density of states near zero energy.  This is a consequence of the short-range correlations, and is also seen in other 1D models \cite{papatsvelik}.  It is natural to assume that such structures are generic, and whenever you have local order, but no long range-order you expect a similar spectrum.  

A physical picture of Fig.~\ref{leefig} comes from noting that there are length and time-scales associated with the fluctuations.  If you examine the system on length-scales smaller than the correlation length, or time-scales smaller than the correlation time, the system appears ordered, and the mean-field theory applies.  Thus if you measure the density of states with an instrument with finite spectral resolution, one can only distinguish the three curves in Fig.~\ref{leefig} if your resolution is sufficiently high.

It is natural to refer to the spectral features in Fig.~\ref{leefig} as a pseudogap. 

%Note, however, that the local order in this example is simply a special case of critical fluctuations

\subsection{Critical Phenomena}
 Any second order phase transition has a critical region near $T_c$ where there is local order.  As one approaches $T_c$ the ordered regions becomes large.  This happens even in conventional superconductors \cite{ar,paraconductivity}, which are not typically thought of as possessing a pseudogap, and one may therefore want to be careful about using that term to describe this phenomenon.  The example in Sec.~\ref{peierls} can be thought of as an extreme example of these critical fluctuations.

\section{Dilute Fermi gases}\label{fg}

In this section we give a quick introduction to the physics of dilute Fermi gases, and explain why they are ideally suited for studying superconducting pseudogaps.  Stajic et al.  were one of the first groups to advocate this perspective \cite{stajic}.

\subsection{Feshbach resonances and the BCS-BEC crossover}\label{crossover}
One of the most beautiful features of nature is its universality.  The physics found in neutron stars at densities of $10^{38}$cm$^{-3}$ and temperatures of $10^{11}K$ can be connected to phenomena in metals ($n\sim 10^{22} {\rm cm}^{-3},T\sim 200 $K) or even ultracold atomic gases ($n\sim 10^{12}{\rm cm}^{-3},T\sim 10^{-9}$K).   A typical Fermi gas experiment involves $N\sim10^6$ atoms of $^6$Li or $^{40}$K, confined by a roughy harmonic optical potential ($V\sim m\omega^2 r^2/2$, with $\omega\sim100$Hz).  Although the nK temperatures are low by absolute standards, the densities are also low.  The temperature scale associated with the density, $T_f\sim \hbar^2 n^{2/3}/m$ is rarely more than 5 times the temperature.  For comparison, in a room temperature metal $T/T_F<0.01$.  %The only reason one can see 

While the phenomena of these disparate systems are very similar, the cold atom experiments are carefully engineered so that the microscopic description is particularly simple.  For example, because of the low densities, only pairwise collisions collisions occur.  Moreover, the temperatures are so low that the 
%
%Atomic gases, with densities of order $10^{14}$cm$^{-3}$ are now routinely cooled to temperatures of 100's of nK or below.  (At least one laboratory has reported a temperature of 500pK \cite{ketterle}.)  At these temperatures the 
DeBroglie wavelength of the atoms are much larger than the range of the interatomic forces.  Under these conditions the details of the interatomic potentials become irrelevant, and scattering can be parameterized by a single quantity, the s-wave scattering length $a_s$ \cite{landauquantum}.  A key feature of cold gas experiments, which make them ideal for studying 
strong-interaction phenomena such as pseudogaps, is that the scattering length can be experimentally tuned 
\cite{tiesinga}.
%\cite{feshbachreview}.

The scattering length is formally defined by considering the phase shift $\delta_k$ between long wavength incoming and outgoing spherical waves: $a_s=\lim_{k\to0} -\delta_k/k$.  This is not a particularly intuitive definition, but
by considering
 a few paradigmatic potentials one can develop an understanding.  First, if one has a smooth potential, one can use the Born approximation,
\begin{equation}\label{born}
\left(a_s\right)_{\rm Born}=\frac{m}{2\pi \hbar^2} \int V(r) d^3 r.
\end{equation}
Thus one associates a positive scattering length with repulsion, and a negative scattering length with attraction.  Moreover, stronger scattering is associated with larger scattering lengths.

A second paradigmatic potential is a hard sphere $V(r)=\infty$ for $r<r_0$ and $V(r)=0$ for $r>r_0$.  For a hard sphere, $a_s=r_0$ is always positive.  One often thinks of the scattering length as the radius of a hard-sphere potential which has the same low-energy scattering properties as the real potential.  Of course, negative scattering lengths do not fit into this paradigm.

Finally it is useful to consider an attractive square well: $V(r)=-V_0$ for $r<r_0$ and $V(r)=0$ for $r>r_0$.  Calculating the scattering length for such a potential is straightforward \cite{feshbachreview}.
%In figure~\ref{attractwell} the scattering length is shown as a function of the well depth.  
For small $V_0$, Eq.~(\ref{born}) holds, and the scattering length is negative.  However, when one makes the
well deeper
the scattering length 
becomes large and negative -- eventually
diverging to $-\infty$ before jumping to $+\infty$.  This jump coincides with the appearance of a bound state in the potential:
 $a_s$ is large and positive if there is a weakly bound state (in which case $E_b=\hbar^2/(m a_s^2)$), and it is large and negative if there is a low energy resonance.  Thus, somewhat counterintuitively, one can have a positive scattering length, even when the potential is attractive.  One way to understand this  behavior is that the scattering states must be orthogonal to the bound state -- and hence a low energy bound state acts similarly to a repulsive potential.

%\begin{figure}
%\includegraphics[width=\columnwidth]{scat.pdf}
%\caption{Scattering length for an attractive square well of radius $r_0$ and depth $V_0$.}
%\label{attractwell}
%\end{figure}

An atomic physicist can modify the scattering potential by applying a magnetic field \cite{feshbachreview}.  The principle is that the atoms can have a bound state whose magnetic moment differs from that of the atoms.  Changing the magnetic field is then analogous to changing the depth of an attractive well.  At a scattering resonance, the bound state becomes degenerate with the scattering state, and the scattering length diverges.  As with the attractive square well example, the scattering length is large and negative when the ``bound state" is slightly above threshold. 
%(of course, by definition it is not a bound state then, but rather a resonance).  
Conversely, the scattering length is large and positive when the bound state is below threshold.  The inverse scattering length $1/a_s$ smoothly crosses zero as one changes the magnetic field.  Nothing dramatic happens at the point $1/a_s=0$.

In the context of Fermi gases,
one refers to the regime where $a_s<0$ as the ``BCS" regime:  When $k_f a_s\ll 0$, the theory of superfluidity developed by Bardeen, Cooper, and Schreiffer applies \cite{bcs}, and the ground state is a superfluid of loosely bound Cooper pairs.  The short-range attraction is the ``glue" holding the particles together.  The regime $a_s>0$ is instead referred to as the ``BEC" regime.  In this regime, there is a two-body bound state, representing a diatomic molecule.  Pairs of atoms combine into these bosonic molecules, which undergo Bose-Einstein Condensation, forming a superfluid \cite{BEC}.  No phase  transition occurs when one changes $a_s$ at zero temperature.  Rather, the size of the pairs just continuously evolves.

For technical reasons, the point $1/a_s=0$ is referred to as ``unitarity" or the ``unitary limit".  An excellent discussion can be found in \cite{feshbachreview}.
%This designation refers to the fact that the only thing limiting the strength of the low energy scattering is conservation  of probability, and the fact that the range of the interactions is extremely small.  This conservation law is encoded in the unitarity of the S-matrix.

The terms ``BCS," ``BEC," and ``unitary," will be used throughout this review.

\subsection{Pseudogap in the BCS-BEC crossover}

Above $T_c$ in the deep BCS limit, the single-fermion excitations will be gapless.  One has a conventional Fermi liquid \cite{fermiliquid}, and it would be surprising if there were any gap-like feature in the single particle spectrum.  The deep BEC limit is very different.  There the normal state consists of a gas of diatomic pairs.  Any single-fermion excitation would require breaking a pair, leading to a spectral gap.  Due to the presence of thermally dissociated molecules, the gap will not be perfect, and there will be an exponentially small density of states at zero energy.  This can be considered a classic example of a pseudogap.

As one moves from $1/a_s=+\infty$ to $1/a_s=-\infty$, the normal state has a smooth crossover from a gas of diatomic pairs with a strong pseudogap, to a normal Fermi liquid, with no gaplike feature.  The pseudogap exists where there are strong short-range pairing correlations, and is absent when the correlations are gone.  Thus pseudogap features continuously grow in strength as one moves towards the BEC side of resonance.  

An important caveat is that if one fits the spectrum to a form like Eq.~(\ref{bcs}), then one expects that in the BEC limit $\mu$ is negative, and $\Delta_s$ is small.  The negative chemical potential encodes the binding energy of the pairs, and the gap is $|\mu|$.  In the BCS limit, one instead expects $\mu$ to be positive, and $\Delta_s$ again small.  It is only in the crossover between these regimes that one expects a fit to a form like Eq.~(\ref{bcs}) will yield significant $\Delta_s$.  Thus, as anticipated in Sec.~\ref{what}, if one defines the pseudogap via such a fitting procedure, then one would say that the the pseudogap can only exist at intermediate coupling.

Ignoring these questions of nomenclature, it is quite difficult to accurately model the gas when $k_f |a_s|\gg1$.  Consequently different theoretical models give different locations and sharpness for the crossover.  This diversity is  seen in comparative studies \cite{chien,levincompare,hucompare,strinatireview,onset,virialspectraltrap} which find
disagreement about the existence/strength of pseudogap spectral features
at the nominal midpoint of the crossover, $1/a_s=0$.  
%As already explained, the controversy is enhanced by competing definitions.

While there is no concensus about the existence of a pseudogap at unitarity, there is agreement that superconducting fluctuations strongly influence the spectrum in this regime.    Additionally, all agree that pairs dominate in the BEC regime.  Thus, at least by the most expansive definitions, sufficiently far in the BEC regime there is a pseudogap.

%
%An interesting observation is that in the deep BEC limit $k_f a_s \gg 0$, the system can be thought of as a gas of bosonic diatomic molecules.  By studying the four-body problem, ?? showed that the interactions between these molecules are described by a repulsive short-range potential with scattering length $a_b=0.6 a_s$ \cite{fourbody}.    Thus as $a_s\to 0^+$, the system is described by a gas of non-interacting bosons.  Conversely, on the BCS side of resonance, as $a_s\to 0^-$ the system is described by a non-interacting gas of fermions.  Thus by changing $a_s$ one tunes from a Bose gas to a Fermi gas.
%

\section{Experimental Probes of  the Normal State of strongly interacting Fermi gases}\label{experiment}

There are a number of ways to experimentally study the pseudogap in Fermi gases.  These range from spectroscopic to thermodynamic.  Here I will describe the main results.  Definitive pseudogap features have been seen in the BEC regime, but the results at unitarity are ambiguous.
%One challenge in the cold gases is that the systems tend to be strongly inhomogeneous:  the center may be superfluid, and the edge normal.  Different experimental techniques have been developed to extract local information.

\subsection{Photoemission spectroscopy}\label{internal}
Photoemission spectroscopy has been one of the most promising probes of cold Fermi gases.
%, and has received immense theoretical analysis \cite{}.   
It takes advantage of the fact that atoms are not simply spin-1/2 Fermions, but have more degrees of freedom.  Typical experiments involve  a mixture of atoms in two collisionally closed hyperfine states, denoted 
$|\!\!\uparrow\rangle$ and 
$|\!\!\downarrow\rangle$.  In photoemission spectroscopy, one drives a transition from the state 
$|\!\!\downarrow\rangle$ to a third states $|x\rangle$.  The experimentalists then look at the number of atoms transfered as a function of the frequency of the drive. 
Reviews of this technique can be found in \cite{levinrfreview,tormareview}.  
%As with many of the other experimental probes, the pseudogap features in the spectroscopic data remain somewhat ambiguous at unitarity.  

A number of different terms are used to describe this technique:
Typical transitions are in the radio or microwave band -- and hence this technique is most often refereed to as ``RF-spectroscopy" or ``Microwave-spectroscopy."  It could also be referred to as ``internal state spectroscopy."  The term  ``photoemission spectroscopy" comes from the fact that a particle  is ``emitted" from the active sector.

Photoemission spectroscopy is easiest to understand when the $|x\rangle$ atoms do not interact with the others.  In that case, the transition rate can be expressed in terms of the single-particle spectral density $G^<(k,\omega)= f(\omega) A_k(\omega)$, which encodes how many ways there are to remove a particle, changing the energy by $\hbar \omega$, and the momentum by $\hbar k$.  The Fermi function $f(\omega)=1/(e^{\beta(\omega-\mu)}+1)$ encodes the filling of the states, and as discussed in section~\ref{what}, $A$ encodes the dispersion.  Repeating our previous example, in an ideal Fermi gas, $A_k(\omega)= 2\pi \delta(\hbar \omega-k^2/2m)$ is non-zero only when $\hbar \omega=k^2/2m$.  Finite lifetime broadens this spectral function.  

In a paired system  one expects $A_k(\omega)$ to have two branches: one associated with removing a particle by breaking up a pair, another with adding an unpaired particle \cite{chinjulienne}.  The existence of these two branches, with a region of low spectral weight between them, is another definition of the pseudogap.  Unfortunately, since it involves removing particles, photoemission spectroscopy is only sensitive to one of the branches.  There have been attempts to ``inject" particles, but so far they have only been applied to the noninteracting system \cite{photoinjection}.
Injection experiments probe $G^<(k,\omega)= (1-f(\omega)) A_k(\omega).$

The connection between photoemission and the spectral function comes from Fermi's Golden Rule,
which encodes conservation of energy and momentum.  In particular, when
 illuminated by light of frequency $\nu$, the rate of production of $x$-state atoms with momentum $p$ is expected to scale as
\begin{equation}
\Gamma_p(\nu) \propto G^<(p,\nu-\nu_0-p^2/2m),
\end{equation}
where $\hbar \nu_0$ is the energy difference between the internal states $|\downarrow\rangle$ and $|x\rangle$ in vacuum, and $\hbar \nu_0+p^2/2m$  corresponds to the energy of the atom emitted in the $|x\rangle$ state.   In solid state experiments, the technique known as ``Angle Resolved Photo-Electron Spectroscopy" measures this same quantity.

Initial experiments \cite{chinrf,greinerrf,bcsbecrf,schuncknature,ketterlerf} averaged over all momentum, measuring
\begin{equation}
\Gamma(\nu)=\int \frac{d^3p}{(2\pi)^2} \Gamma_p(\nu).
\end{equation}
Later experiments developed techniques to directly observe $\Gamma_p(\omega)$ \cite{momentumresolved}.  The resulting data can be deconvolved to produce $A(k,\omega)$.  One can then analyze this quantity for signs of pairing.  Because of the Fermi occupation factor, one only has access to freqencies below the chemical potential.

Some of the first evidence for pairing in the BCS-BEC crossover came from such experiments.  Chin et al. found that at high temperatures the absorption spectrum was sharply peaked, $\Gamma(\nu)\approx \delta(\nu-\nu_0)$  \cite{chinrf}.
This behavior would be expected for a non-interacting gas, and is indicative of a lack of pairing.
  At low temperatures, they instead saw a broad peak, which vanished below a threshold $\nu_{\rm th}>\nu_0$.  The difference $\delta\nu=\nu_{\rm th}-\nu_0$ was taken as a measure of the gap (though at the time competing theories were put forth for the observations which did not require superfluidity \cite{massignan,generic}).  
At intermediate temperatures they observed a bimodal distribution -- formed from a sharp peak at $\nu_0$, and a broader peaked centered at higher frequencies.  This structure naturally arises  from the inhomogeneity of the cloud:  The weakly interacting wings contribute a delta-function at $\nu_0$, while averaging over the more strongly interacting central region yields the broad peak \cite{kinnunen,regal,schunck,ohashi,yanhe,generic,massignan}.  Variants of this technique were developed in several other labs, and gave strong evidence for pairing at low temperatures \cite{greinerrf}. 

Early experiments were complicated by trap inhomogeneities.  At MIT they developed a spatial resolved spectroscopy \cite{mitspatialresolve}.
%, which was largely incompatible with the more powerful momentum-resolved techniques \cite{momentumresolved}.  
More recent experiments at JILA implemented a hybrid method which probes only the center of the cloud, but is momentum resolved \cite{realandk,localPES}.  This position and momentum momentum resolved photoemission spectroscopy is one of the most important tools that has been developed by the cold gas community.

A second important technical issue involved final state interactions.  The assumption that the final state atoms are non-interacting is not always valid \cite{strinatiRFAL,final,schuncknature}.  These final-state interactions are particularly problematic in $^6$Li, where all collision channels have nearby Feshbach resonances.  Fortuitously, the final-state interactions are weak in $^{40}$K, and the most quantitative experiments involve those atoms.  Note, that even with final-state interactions, one can learn a great deal about pairing from internal state spectroscopy, but the analysis is more complicated, and requires modeling.

Since the primary technical issues have been resolved, there is now excellent spectroscopic data \cite{localPES,realandk}, especially at unitarity, and one can reasonably ask if the experimentally observed $A_k(\omega)$ is indicative of a pseudogap.  Unfortunately, there is no simple answer to this question.  The primary difficulty is that 
at unitarity the normal-state spectra are broad and indistinct.  Such broad spectral functions are expected, and are indicative of the short quasiparticle lifetimes in the strongly interacting gas \cite{lifetime,palestinilifetimes}.  These short lifetimes are consistent with the hydrodynamic behavior of the gas \cite{hydro}.  Adding to the difficulty is the fact that because one is extracting particles, the experiment only provides $A_k(\omega)$ for energies below the Fermi level.  

The end result is that despite attempts to argue one way or another \cite{localPES,lifetime}, the photoemission spectroscopy experiments are ambiguous about the presence of a pseudogap at unitarity.  As previously explained, such ambiguity is neither surprising, nor alarming.  In tuning from the BEC to BCS regime, one expects a smooth crossover between a normal state dominated by bosonic pairs, to one dominated by free fermions.  The former will show a distinct pseudogap, while the latter will have absolutely no pseudogap features.  Unitarity sits between these limits, and the spectrum is appropriately ambiguous.

\subsection{Equation of State}\label{thermo}
The second major probe of atoms in the BCS-BEC crossover is thermodynamics \cite{ketterleeos,parisEOS,MITEOS,thermo,strinatiEOS,compress}.  The most detailed studies have concentrated on the equation of state relating the pressure $P$ to the temperature $T$ and the chemical potential $\mu$ \cite{parisEOS,MITEOS} -- finding remarkable agreement between numerical Monte-Carlo techniques, and experiments.  

In principle the equation of state can distinguish between a Bose and Fermi gas -- and hence can identify signatures of pairs in the normal state.  For example, as $T\to 0$ in an ideal Fermi gas, the pressure varies quadratically with temperature, $P\to P_0+\lambda T^2$, where $P_0$ and $\lambda$ are constants related to the density.  Fermi liquid theory predicts that this temperature dependence is robust against interactions \cite{fermiliquid}.  The equation of state of an ideal Bose gas is very different.  If one stipulates that $T\gg T_c$, then the Bose gas obeys the ideal gas law and  the pressure vanishes linearly with $T$.  One might expect that a pseudogap state would follow the bosonic prediction.  One should be cautious, however: In the pseudogap regime, the atoms are degenerate, and interactions are strong.  Thus it would be naive to expect that the equation of state of the psuedogap obeys an ideal gas law.

Experiments \cite{parisEOS,MITEOS}, and theory \cite{MITEOS,strinatiEOS,levinEOS}, find that in the normal state of the unitary Fermi gas ($1/k_f a=0$), the equation of state is well fit by the empirical curve $P=P_0+\lambda T^2,$ as predicted by Fermi liquid theory.  The Paris group went so far as to state that this observation is {\em incompatible} with a pseudogap \cite{parisEOS}.   This conclusion is not universally accepted, as there are many pseudogap theories which predict this same temperature dependence
\cite{levinFL,levinEOS,chienlevin}.
%and the experimental observations are compatible with theories that produce a 
%spectral pseudogaps \cite{levinEOS}.   
Regardless, the experimental results certainly show that structureless noninteracting bosonic pairs are not dominating the physics.  It would be extremely interesting to see how the equation of state evolves as one tunes towards the BEC limit, where pairs are more tightly bound, and pseudogap features are expected to become stronger.   If there are deviations from the Fermi-liquid predictions, they will be most obvious there.

The techniques used to measure the equation of state are quite elegant.  They take advantage of the fact that the trapped clouds are in hydrostatic equilibrium:  Consider a small cube of gas at position $r$, with volume $d^3r$.  The trapping force in the $\hat x$ direction is  $F_{\rm trap}=-n d^3r \partial_x V(r)$.  This should be balanced by the hydrodynamic forces from the neighboring regions, $F_{\rm hydro}=  d^2r P({\bf r}-\hat x dr/2)- d^2r P({\bf r}+\hat x dr/2)=d^3r \partial_x P(r)$.  Experimentalists know $V({\bf r})$, and measure $n({\bf r})$, and integrate the hydrostatic equations, $F_{\rm trap}+F_{\rm hydro}=0$, to find $P({\bf r})$.   They thereby parametrically produce a relationship between the density and pressure.  One can simplify the analysis, and reduce systematic errors, by appropriately engineering the trapping potential \cite{zwierleinhomo}.

There are a number of other techniques which allow one to access thermodynamic quantities \cite{thermo,fluc,compress,criticalvel,mitimage,lingham}.  These include looking at the response of the cloud to a perturbation \cite{lingham,compress,criticalvel,bragg,bruun} and measuring local fluctuations \cite{fluc}.
Such studies have been useful for judging the quantitative accuracy of competing theories, and have helped us develop a phenomonological understanding of both the superfluid and normal state.

\subsection{Spin Susceptibility}

As argued by Trivedi and Renderia in the context of superconductors \cite{trivediranderia}, 
a more direct thermodynamic probe of normal-state pairing is spin susceptibility.  If all of the fermions are bound into pairs, then the gas should be strongly diamagnetic.  Thus one expects that the $k=0$ spin susceptibility of the pseudogap state should be suppressed.  Indeed, Monte-Carlo calculations find a dramatic drop in this susceptibility as one moves from the BCS to BEC side of resonance \cite{exptsusc}.   Diagramatic theories find a similar suppression \cite{ohashispin}.  At unitarity, experiments find that the spin susceptibility is smaller than one would find for a non-interacting gas \cite{exptsusc,mitsusc}, in agreement with Monte-Carlo calculations \cite{mcdip}.
%Note, the Monte-Carlo calculations suggest that the diagrammatic theories underestimate the suppression.

Several different techniques have been used for measuring the spin susceptability.  For example, at MIT they measured  density profiles in magnetic field gradients \cite{mitsusc}.  Using arguments analogous to those in section~\ref{thermo}, they could then  relate the spatial variation of the polarization to the susceptibility.  Alternatively, in Paris they measured the equation of state at finite spin polarization
\cite{exptsusc}.  The susceptibility was then  calculated through a numerical derivative.  

As with the spectroscopic data, there is some controversy about the signatures of the pseudogap in spin susceptibility.  For example,
 Nascimb\`ene et al. argued that instead of a suppression in the susceptibility, the pseudogap state should be characterized by a non-linear susceptibility \cite{exptsusc}: they argue that the susceptibility should be suppressed for small fields, but restored at larger fields.  The idea being that a strong enough field will break up the pairs, restoring normal Fermi-liquid behavior.  Both experiments, and Monte-Carlo calculations, fail to find this non-linearity near unitarity.  Thus Nascimb\`ene et al. conclude that the normal gas at Unitarity does not possess a pseudogap, and hence the pseudogap regime is restricted to the BEC side of resonance. 
 
Tajima et al. provide an alternative picture \cite{ohashispin}.  They used a T-matrix approximation to calculate the susceptibility as a function of temperature throughout the BEC-BCS crossover.  Tajima et al. found that the susceptibility is a non-monotonic function of temperature, with a peak at $T=T_s$.  They interpret $k_B T_s$ as a pairing energy:  for $T>T_s$, pairs are unimportant, while for $T<T_s$ there are fewer and fewer free spins.  They refer to the region $T_c<T<T_s$ as the ``spin-gap" regime -- a phrase which is somewhat more precise that ``pseudogap."   At unitarity, their approximations yield $T_s/T_c\sim 1.7$.  Although there have not yet been any experiments that have measured $T_s$, the techniques in \cite{exptsusc} or \cite{mitsusc} could be used for such a study.
Wulin et al. \cite{levinspin} used a different T-matrix theory to calcuate the spin susceptability.  They obseved a low-temperature suppression of the spin susceptibility, but did not study the peak.

Tajima et al. also explored the relationship between their susceptibility and the density of states at the Fermi surface.  They compared $T_s$ with two other scales $T^*$ and $\tilde T^*$:  $T^*$ is the temperature at which the density of states at the Fermi energy is maximal and $\tilde T^*$ is the highest temperature at which a depression in the density of states first appears at $T_c$.  They find that within their theory, $T_s\sim T^*\gg \tilde T^*$.  Moreover, at unitarity, $\tilde T^*\sim T_c$.  Thus, even within a single theory, the different  facets of pairing turn-on at different scales.  As already mentioned, other authors use different notation.

One can gain even more information about the quantum state by studying the dynamical spin response \cite{dynspin,dynspintheory,dynspin2}.  There does not, however, appear to be any simple story connecting these dynamical experiments to normal-state pairing.

\subsection{Transport}
Closely related to the thermodynamic probes, are transport measurements.  Although a mainstay of solid state, transport is harder to access in cold atoms.  Transport is also often harder to interpret.  These experiments have given profound insight into the nature of the normal state  of the unitary Fermi gas, but no  clear connection has been made to pseudogaps \cite{transportreview}.

The most basic transport-like experiment is time-of-flight expansion.  One simply turns off all trapping potentials, and allows the cloud to free-fall under gravity.  As it falls, the cloud expands, and by observing this expansion dynamics one can infer its properties.   In particular, early experiments on anisotropic clouds found that the aspect ratio of the cloud reversed during expansion, a sign of strong interactions with connections to observations in heavy ion coliders \cite{anisoexp}.  Further information came from exciting collective modes of the cloud \cite{colmodes,hydro}, and additional expansion experiments \cite{expansion,universal}.  Such experiments clearly showed that the strongly interacting Fermi gas behaves hydrodynamically.  Later experiments extracted transport coefficients, such as viscosity \cite{universal,damping}.

More relevant to normal-state pairing are studies of spin transport.  There have been a number of measurements of spin diffusion based upon either studying the dynamics of the magnetization of a cloud in a field gradient \cite{thywissenspin}, or the relaxation of imposed spin textures \cite{mitsusc,zwierleinspin, othertransport}.  There has also been a number of studies of spin waves in the more weakly interacting limit \cite{spincurrents}.  Wulin et al. argue that one distinct signature of pseudogap physics is the relationship between the spin susceptibility and spin diffusivity \cite{levinspin}.  In particular, they argue that the experiments in \cite{othertransport} imply a pseudogap.   

Another class of experiments attempts to replicate a more traditional transport geometry by producing a dumbell shaped trap with two large reservoirs connected by a narrow channel \cite{junction}.  The experimentalists created an initial population imbalance between the two reservoirs, then watched the subsequent dynamics.  In the non-interacting limit they observed quantized conductance, as predicted by the Landauer formalism \cite{giamarchi}.  As they increased the strength of (attractive) interactions, they observed a significant enhancement of the conductance.  The leading theories suggest that this enhancement is due to pairing  -- possibly the formation of superfluid regions \cite{demlerpair} or normal-state pairing \cite{uedapair,zhangpair}.   As such, these measurements may be indicative of pseudogap physics.

\subsection{Contact}

In addition to these probes, which have analogs in condensed matter physics, there are a number of observables which are unique to cold atoms -- which can be used to learn about the presence of normal-state pairing.  The most well-studied of these is the ``contact,"  introduced by Tan \cite{tan}, and explored by many other authors \cite{othercontact}.  As its name suggests, the contact is a measure of short-range correlations between particles.  Remarkably, in systems with short range interactions, all short-distance (high energy) physics is parametrized by a single number $C$.  For example, for large $k$ the momentum occupation factor $n(k)\to C/k^4$; for large frequencies, the radio frequency absorption spectrum (sec.~\ref{internal}) is $I(\omega)\to C/(2^{3/2} \pi^2 \omega^{3/2})$; and the  thermodynamic derivative of the free energy with respect to the scattering length is
\begin{equation}
\frac{\partial F}{\partial a^{-1}}=-\frac{\hbar}{4\pi m} C.
\end{equation}
Larger contact implies a greater likelyhood for two particles to be close together.  Thus there is a connection between the contact and pairing.

Pieri et al. \cite{strinatiRFAL} advocated identifying, $C=(m \Delta_\infty)^2$, where they take $\Delta_\infty$ to be a local measure of pairing.  
Indeed, the contact in a BCS superfluid has this form with $\Delta_\infty$ replaced with the superfluid gap.  Moreover, they argued that this identification was consistent with models of normal-state pairing.  One should be cautious, as there are numerous physical settings where it is not natural to identify short range correlations with pairing. 

As with other cold-atom probes, inhomogeneous broadening can complicate the extraction of the contact.  Moreover, there are particular technical issues associated with the different ways of measuring the contact.  For example, one can in principle extract the contact from time-of-flight expansion images \cite{jintof}.  Interactions during time-of-flight can, however, render these results unreliable.  Similarly, the spectroscopic probes of contact must deal with final-state interactions.  

In one of the first theoretical calculation of the normal state contact, Palestini et al. \cite{palestinicontact} saw that at $T_c$, the contact monotonically decreased as one moved from the BEC to BCS limit, with a relatively sharp crossover in behavior near $k_f a_s=-2$.  They also explored the temperature dependence of the contact, finding a slow rise as temperature is lowered, with a sharper rise near $T_c$.  One can argue that this region of enhanced $C$ near $T_c$ is related to normal-state pairing.  Other theoretical approaches yielded similar results \cite{bdmcontact,goulko,balcontact}, though there are disagreements about quantitative details.

Several experimental techniques have been used to study the contact.  Kuhnle et al. used Bragg spectroscopy to measure the trap-averaged contact in a  unitary Fermi gas \cite{kuhnle}.  They observed the expected monotonic decrease in the contact as a function of temperature, but due to trap averaging their results were not particularly discriminating between competing theories.   Later
Sagi et al. extracted the contact from the high energy tail in photoemission spectroscopy \cite{jincontact2,jincontact}, using slicing techniques to isolate a roughly homogeneous region \cite{localPES}.

%\subsection{Mesoscopic Junctions}
%Although not designed to explore pseudogap physics, there has been some speculation that transport experiments in Zurich \cite{} have seen evidence of normal-state pairs in a strongly interacting Fermi gas \cite{giamarchi}.
%
%An alternative explanation of the data involves the formation of superfluid regions due to a confinement induced resonance \cite{demleresslinger}
%.

\subsection{Photoassociation}
Another probe that is unique to cold atoms is photoassociation \cite{photoassociation,photoassociationreview}.  There one drives a transition between a scattering state and a bound state.  The rate of transition is proportional to the overlap between the quantum states, giving a measure of the short range pair correlations.   
Varients of this technique were used extensively to study pair correlations in thermal gases \cite{thermalphoto}, and degenerate Bose gases \cite{bosephoto}, and form the basis of modern techniques for producing ultracold molecules \cite{ultramol}.  

Experimentalists at Rice University studied the low temperature photoassociation signal throughout the
 BEC-BCS crossover \cite{fermiphoto}.  Through this measurement they quantified the pairing correlations in the ordered state.  Unfortunately, there have been no systematic studies of the photoassociation signal in the normal state of the BCS-BEC crossover.  Such measurements would be useful for quantifying precursors of pairing.
 
 A closely related probe is inelastic two and three body collisions.  Du et al. measured the rates of these processes, arguing that in the pseudogap regime they are dominated by collisions between atoms and pairs \cite{inelastic}.

\subsection{Quench}
The last probe that I will discuss  is the response of the gas to a sudden change in the scattering length.  In particular, several groups have extracted useful information by rapidly sweeping from the BCS to BEC side of resonance \cite{sweep}.  The idea is that the loosely bound BCS pairs will be projected into tightly bound dimers.  These dimers are then robust enough to directly measure.  This technique has been used to detect a BCS condensate \cite{sweep}, vortices \cite{vortsweep}, and dark solitons \cite{solitonsweep}.  In the normal state one might be able to use such a sweep to get information about the momentum distribution of normal-state pairs.  Unfortunately, modeling the sweep is relatively complicated \cite{sweepmodel,sweepmodel2}, and it is hard to extract quantitative data from these experiments.

\section{Theoretical Models of the Normal State of strongly interacting Fermi gases}\label{theory}
Here I present the models and calculations which have investigated the normal state properties of strongly interacting Fermions in the BCS-BEC crossover.  The focus will be on developing an intuition about the approaches.  For technical details, readers will be directed to the primary litterature, and other review articles.

The theory of the BEC-BCS crossover predates cold atom experiments.   It grew out of chain of research introduced by Eagles \cite{eagles}, Leggett \cite{leggett}, and Nozieres and Schmidt-Rink \cite{nozieres} -- which was further developed by Sa de Melo, Randeria and Engelbrecht \cite{earlyranderia}, then taken up by a larger community \cite{chen,micnas,ranninger}.  

As already explained in the introduction, modeling a correlated liquid is difficult.  Neither a dilute gas, nor a solid are good starting points.  We are interested in a normal state, for which there is no long-range order to expand about.  A number of approaches have been developed:  scaling theories, Monte-Carlo methods, high temperature expansions, higher order perturbation theories, and self-consistent diagrammatic expansions.  Each of these yield their own insight into the problem.

\subsection{Scaling Theories}\label{scaling}
Nearly all attempts to understanding the strongly interacting normal state makes use of the relatively small number of dimensional quantities in this system: the Fermi energy $E_f=k_f^2/2m$, the scattering length $a_s$, and the temperature $T$.  At unitarity, $a_s=\infty$, and there are at most two scales in the system.  Thus all thermodynamic functions take simple forms.  For example, the pressure can be written as $P= E_f k_f^3 f(\beta E_f)$, where $\beta=1/k_B T$ and $f$ is a function of one variable \cite{bertsch}.  This scaling also leads to a number of relations between thermodynamic functions \cite{castin}, and can also be applied to dynamical quantities.  Any theory of the pseudogap needs to obey these relations.  The techniques discussed in section \ref{mc} through \ref{diag} can be used to calculate or constrain the scaling function.

\subsection{Monte-Carlo Methods}\label{mc}

A number of stochastic methods have been developed for studying the properties of strongly interacting Fermi gases.  Typically they give access to thermodynamics, and can be compared to the experiments 
 described in section~\ref{thermo}.
Initial stochastic approaches to understanding the strongly interacting Fermi gas focussed on ground-state properties, using a number of Monte-Carlo techniques \cite{carlsonzeroT, trentozeroT}.
One important feature of the attractive Fermi gas is that there exist Monte-Carlo approaches without ``sign problems" \cite{nosign}, and hence, with enough computer power, stochastic methods can produce results with arbitrary accuracy.

A large number of different stochastic techniques have been applied to the finite temperature gas.  The earliest of these are the Auxilliary-Field Monte-Carlo calculations of Bulgac, Drut and Magieski \cite{bulgac}, the hybrid Monte-Carlo calculations of Wingate \cite{wingate} and of Lee and Sch\"afer \cite{leeschaefer}, the determinant Monte-Carlo calculations of Burovski, Prokof'ev, Svistunov, and Troyer \cite{determinantmc}, and the path integral Monte-Carlo calculations of Akkineni, Ceperley, and Trivedi \cite{trivedi}.  The fact that these very different formalisms agree well with one-another is an excellent indicator of the accuracy of the modeling.

Monte Carlo techniques typically map the thermodynamics of a quantum system onto the thermodynamics of a classical system.  One can then sample the classical distribution to calculate thermodynamic properties of the quantum system.
  The quantity most carefully compared with experiments is the equation of state.  For example, in 2012, Van Houcke et al.  presented a careful comparision of ``Bold-diagramatic Monte Carlo" with experiments at MIT \cite{amherstMIT}.   Data from the same experimental group \cite{ku} was then later compared with both hybrid and auxiliary field Monte Carlo approaches \cite{mceoscompare}.  Subsequent studies continued to expand the parameter range and accuracy of the numerical calculations
\cite{oldgoulko,goulko,amherst,pollet,exptsusc,lmc,af}, as well as calculating other observables, such as contact \cite{contactmc}.  From these comparisons, there is no doubt that these ab-initio theories accurately model the Fermi gas throughout the BCS-BEC crossover.  Unfortunately, as discussed in section~\ref{thermo}, these thermodynamic results are somewhat ambiguous about the presence of normal-state pairing.  

Recently progress has been made in extracting spectral data from 
 Monte-Carlo calculations \cite{realtime, polaronspectrum, magierski,magierskigap,wlazlowski,sheehy,mcdip}.  For example, Magierski et al. used numerical analytic continuation to extract single particle
spectral densities from Monte-Carlo data \cite{magierski}.  Remarkably, they find a distinct gap-like feature in the spectrum at unitarity.  
Not only does their density of states show a dramatic dip \cite{mcdip} near $T_c$, but the spectral density $A(k,\omega)$ shows two distinct branches which are well separated from one-another.   From signatures like these, Magierski et al.  argue that the criterion for observing a pseudogap is $1/k_f a_s >-0.05$ \cite{onset}. 
Similar results are seen in dynamical cluster quantum Monte Carlo \cite{sheehy}.
The ultimate reliability of these calculations needs to be confirmed by appropriate extrapolation to infinite system size, and infinitesimal lattice spacing.

\subsection{High Temperature Expansions}
One simple limit of any system is ``infinite temperature,"  where all states are equally likely, and thermodynamic properties are trivial.  Systematic corrections can be calculated in a power series in the fugacity $z=e^{\beta\mu}$, where $\beta=1/k_B T$, and the chemical potential $\mu$ is negative.  The resulting expansion for the free energy is known as the {\em virial expansion}, and has been calculated to second \cite{virialsecond},  third \cite{virialthird,hightemp} and fourth \cite{virialfourth,ngampruetikorn} order in $z$.  The third order series agrees quantitatively with experiments for $k_B T/E_f>0.8$.  After applying appropriate resummation techniques, it also agrees qualitatively for $k_B T/E_f>0.2$.  Consequently, one only expects that the high temperature expansion to yield information about the pseudogap on the BEC side of resonance, where this phenomenon persists to high temperatures.  Beyond this direct information, the high temperature expansion is useful for calibrating other theories, as it is well-controlled.  A thorough review of the subject can be found in \cite{virialreview}.

In addition to the equation of state, high temperature expansions yield the contact \cite{virialcontact} and response functions \cite{sunleyronas,virialspectraltrap,shenresponse,ngampruetikorn,virialstructure}.  For example, in \cite{virialspectraltrap}, Hu, Liu, Drumond, and Dong, used a high temperature expansion to calculate the spectral density of a trapped gas at $k_B T/E_F=0.7$.  They saw a distinct gap-like feature on the BEC side of resonance ($k_f a_s=1$), but found that the spectra at $1/(k_f a_s)=0,-1$ were broad without clear indications of pairing.   Similar results were found by Ngampruetikorn, Parish, and Levinson \cite{ngampruetikorn}.

%\subsection{Higher Order Perturbation Theories}

\subsection{Dimensional Expansions}
As reviewed in \cite{wilsonkogut}, there is an ingenious approach to statistical mechanics where one treats the dimension of space $d$ as a continuous variable.  If one can solve the problem for a particular $d=d_c$, one can then use perturbation theory in $\epsilon=d-d_c$ to learn about the physics in other dimensions.

In 2006, Nishida and Son applied this technique to the strongly interacting Fermi gas \cite{epsilon}, taking advantage of an result from  Nussinov and Nussinov \cite{nussinov}, which showed that the problem simplifies in $d=4$ and $d=2$.  In low dimensions ($d\leq 2$), an arbitrarily weak attraction leads to a bound state, providing a mapping between the unitary limit (where a bound state is at threshold) and the non-interacting limit.  In high dimensions ($d \geq 4$), bound states have an extremely large weight at the origin, suggesting a description of the unitary gas in terms of pointlike bosons.

 Nishida and Son first expanded about $d=4$ to calculate the zero temperature properties of the dispersion and the equation of state \cite{epsilon} at unitarity. Later, they considered the expansion about $d=2$ \cite{epsilon2}.  By combining these two expansions they were about to bound the properties of the unitary gas in $d=3$.  Nishida extended these results to finite temperature \cite{nishida}.  He found that near $T_c$ both Fermionic and Bosonic degrees of freedom were important for the thermodynamics -- a feature naturally interpreted in terms of normal state pairs.  Other authors explored the BCS-BEC crossover \cite{cheneps}, and carried the expansion to higher order \cite{arnold,moreeps}.

\subsection{T-Matrix Approaches}\label{diag}
The first theory of the normal state in the BCS-BEC crossover was developed by Nozieres and Schmidt-Rink (NSR)\cite{nozieres}.  It was framed as a ``diagramatic" theory, where one ressums infinite sets of diagrams which represent terms in a perturbative expansion.  Nozieres and Schmidt-Rink's approach 
% summed a set of diagrams which captured the underlying physics.  Their approach 
roughly corresponded to exactly solving the two-body problem, accounting for the presence of all the other particles through Pauli blocking.  The importance and influence of this work cannot be understated.  It, for example, provides a model for how the superfluid transition temperature evolves in the crossover.   
Variants of the NSR theory have been one of the key analytic approaches to superconductivity \cite{km}.
In the 1990's, Sa de Melo, Randeria, and Engelbrecht reformulated the NSR theory, and elucidated both the phase diagram, and the excitations \cite{bcsbecearly1}.

There have been numerous attempts to either justify or improve on the NSR theory, resulting in a rich set of ``T-Matrix" approaches, so-named because the class of diagrams summed \cite{levincompare,hu2,hu3,luhu,earlyhaussmann,laterhaussmann,Tsuchiya,stajic,kinnunenhartree,pierit,micnast,gubelstoof,pieristrinati,chen2,bcsbecpseudo,haussmann,stajic,strinati,perali}.  The three main T-Matrix theories are reviewed and compared in a number of articles by Chen and collaborators \cite{chen,levinxx,levincompare,hucompare}.  As previously explained, the theories predict different strengths of gap-like features at unitarity, and there is active debate about which of these features should be considered indicative of a pseudogap.  
In all models, the gap-like features become stronger as one moves towards the BEC limit, and weaker as one moves towards the BCS limit. 
%, but all agree that 
%there is a strong pseudogap in the deep BEC limit, and none in the deep BCS limit.  
%There were several years during which these were the only theories of the normal state, and as such they have important historical importance.  
%They are valuable as semi-analytic approaches in which one can easily add or subtract various physical processes, and readily produce qualitative narratives.  
%When compared with quantum Monte-Carlo, it appears that
%T-matrix approaches systematically underestimate the strength of pseudogap features in the 
%single-particle spectrum \cite{onset, magierski,haussmann}, with the NSR approach showing the weakest pseudogap.  The Monte-Carlo calculations also predict a larger suppression of the spin-susceptibility.  
%While their quantitative accuracy is a matter of debate, t
There are ongoing 
 attempts to reformulate these theories in ever-more accurate ways  \cite{mulkerin}.

\subsection{Other Approaches}
Sections~\ref{scaling} through \ref{diag} detailed the most popular theoretical approaches to calculating properties of the normal state in the BEC-BCS crossover.  There are, however, several other techniques.  For example, a number of authors introduced ``large $N$ expansions'' in which they replace the spin-1/2 fermions with spin $N/2$ particles.  In the limit $N\to\infty$ one can make simplifying assumptions, or derive renormalization group equations \cite{largeN,sachdevlargeN}.  There are also theories based
 upon the 2-body S-matrix \cite{howleclair}, Brueckner-Goldstone theory \cite{bg}, Dyson-Schwinger equations \cite{ds}, Bethe-Salpeter equations \cite{bs}, projected wavefunctions \cite{cazalilla}, renormalization groups \cite{gubbelsrg,rg,sachdevlargeN,morerg,brg},
 operator product expansions \cite{opepseudo},
and effective field theories \cite{effective}.  A number of authors have attempted to make simplified theories which have Hartree-like structure \cite{weiler,kinnunenhartree}.  Generically, each of these techniques is designed to highlight some facet of the phenomonology, and where they are reliable they agree with one-another.  They represent a
wonderful toolbox for understanding this rich system.
%ir relative merits are mainly of interest to experts in the field.

\section{Outlook}\label{summary}
Throughout this review I have tried to present a relatively simple story:  In the deep BEC regime the normal state is described as a gas of weakly interacting bosonic pairs, in the deep BCS regime the normal state is described as a gas of weakly interacting fermions, and the phenomena in the intermediate regime crossover region shows facets of both descriptions.  Some of these facets are traditionally described as signatures of a "pseudogap".  In the introduction, I argued that this phenomonology is an example of a larger principle,  and that pseudogap phenomena are generic features of strongly interacting fermions.  In this section I would like to very briefly reiterate that viewpoint, touch on important questions, and explore the outlook for this area of study.

I should first note that there are obvious examples of strongly interacting fermions which are unlikely to have a pseudogap description.  For example, the metal-Mott crossover involves very different physics \cite{mottmetal}.  I should also note that the example in Sec.~\ref{peierls} shows that one can have a pseudogap without pairing.  Despite these limitations, I still contend that the pseudogap in the BEC-BCS crossover provides an important paradigm, beyond its application to cold atoms.  It is generically true that near a phase transition, the unordered state shows precursors of the ordering.  What is special about the cold atom system is that one has  a control parameter ($1/k_f a$), which allows one to tune the strength of the incipient order.  
%The other important lesson from the BEC-BCS crossover is that the phenomenology of superfluid fluctuations can be complex.  There are multiple manifestations of these fluctuations, and they can become relevant at different temperatures.

Is the BCS-BEC crossover relevant to strong-coupling superconductors?  As already discussed, the physics of the cuprates appears to involve some non-superconducting order.  Nevertheless, one must acknowledge that the pairing energy scale is large in those materials, and near $T_c$ there must be precursors of pairing.  Disentangling those precursors from the other phenomena is a difficult task.

Cold atoms in the BCS-BEC crossover are an incredibly fascinating system.  The pseudogap physics described here is just one facet of them.  There has been remarkable developments in lower dimensions \cite{twod,oned} and in investigation of spin polarized gases \cite{imbalanced}.  There is great interest in exploring these systems on lattices, where one hopes to make closer connections to solid state systems.  One extremely promising recent development is the construction of ``Fermi gas microscopes," in which one can detect the location of every particle in the gas \cite{fgm}.  Other exciting directions include: p-wave interactions \cite{pwave}, pairing with more spin components \cite{higherspin}, and longer range interactions \cite{longrange}.

\section*{Acknowledgements}
I would like to thank Cheng Chin, Jami Kinnunen, Kathy Levin, Nikolay Prokof'ev, Henk Stoof, and Martin Zwierlen for critical comments.  This material was based upon work supported by the National Science Foundation under Grant No. PHY-1508300.


\begin{thebibliography}{55}


\bibitem{chevymora}
Frederic Chevy and Christophe Mora, Ultracold Polarized Fermi Gases,
Rep. Prog. Phys. {\bf 73}, 112401 (2010).



\bibitem{compilations1}
Zwerger, The BCS-BEC Crossover and the Unitary Fermi Gas, (Springer, Heidelberg, 2011).




\bibitem{tormareview}
P\"aivi T\"orm\"a, Physics of ultracold Fermi gases revealed by spectroscopies, 
Physica Scripta, {\bf 91}, 043006 (2016);
P\"aivi T\"orm\"a, Spectroscopies -- theory, in 
P\"aivi T\"orm\"a and Klaus Sengstock,
{\em Quantum Gas Experiments -- Exploring Many-Body States} 
(Imperial College Press, London, 2015) pp 199-250.

\bibitem{BCSBECreview}
Mohit Randeria, and Edward Taylor, 
BCS-BEC Crossover and the Unitary Fermi Gas,
%	Annual Review of Condensed Matter Physics, 
	Ann. Rev. Cond. Mat. Phys.
	{\bf 5}, 209 %-232 
	(2014).


\bibitem{jinreview}
C. A. Regal and D. S. Jin, 
Experimental realization of the BCSÐBEC crossover with a Fermi gas of atoms, 
Adv. At. Mol. Opt. Phys. {\bf 54}, 1
(2006).

\bibitem{ketterlereview}
W. Ketterle and M.W. Zwierlein, 
Making, probing and understanding ultracold Fermi gases. 
in Ultracold Fermi Gases,
in Proceedings of the
International School of Physics ÔÔEnrico Fermi,ÕÕ Course
CLXIV, edited by M. Inguscio, W. Ketterle, and C.
Salomon (IOS, Amsterdam, 2008).


\bibitem{compilations2}
M. Inguscio, W. Ketterle, and C. Salomon, {\em Ultra-cold Fermi Gases: Varenna on Lake Como, Villa Monastero, 20-30 June 2006} (IOS Press, Amsterdam,  2007).

\bibitem{compilations3}
W. Ketterle and M.W. Zwierlein. Making, probing and understanding ultracold Fermi gases, in M. Inguscio, W. Ketterle, and C. Salomon, {\em Ultra-cold Fermi Gases: Varenna on Lake Como, Villa Monastero, 20-30 June 2006} (IOS Press, Amsterdam,  2007).


\bibitem{bloch}
I. Bloch, J. Dalibard, and W. Zwerger, Many-body Physics with Ultracold Gases,
Rev. Mod. Phys. {\bf 80}, 885 (2008).


\bibitem{feshbachreview}
Cheng Chin, Rudolf Grimm, Paul Julienne, and Eite Tiesinga, Feshbach Resonances in Ultracold gases,
Rev. Mod. Phys. {\bf 82}, 1225 (2010). 

\bibitem{giorginireview}
S. Giorgini, L. P. Pitaevskii, and S. Stringari. Theory of ultracold atomic Fermi gases. Rev. Mod. Phys., {\bf 80}, 1215  (2008).




\bibitem{levinhulet}
K. Levin, Randall G. Hulet,
The Fermi Gases and Superfluids: Short Review of Experiment and Theory for Condensed Matter Physicists, in
Kathryn Levin, Alexander Fetter and  Dan M. Stamper-Kurn, {\em Contemporary Concepts of Condensed Matter Science}, (Elsevier, Oxford, 2012).


\bibitem{melo}
Carlos A. R. S\'a de Melo, 
When fermions become bosons: Pairing in ultracold gases,
Physics Today {\bf 61}, 45 (2008).

\bibitem{chen}
Q. Chen and J. Wang, 
Pseudogap phenomena in ultracold atomic Fermi gases,
Frontiers of Physics, {\bf 9}, 539 (2014).

\bibitem{levinxx}
Qijin Chen, Jelena Stajic, Shina Tan, and K. Levin,
BCS-BEC crossover: From high temperature superconductors to ultracold superfluids.
Physics Reports, {\bf 412}, 1 (2005).


\bibitem{twod}
Jesper Levinsen, Meera M. Parish,
Strongly interacting two-dimensional Fermi gases,
in Kirk W Madison, Kai Bongs, Lincoln D Carr, Ana Maria Rey, and Hui Zhai,
{\em Annual Review of Cold Atoms and Molecules, Volume 3} (World Scientific, 2015), pp. 1-75.
%Andrea M. Fischer, Meera M. Parish,
%Quasi-two-dimensional Fermi gases at finite temperature,
%Phys. Rev. B 90, 214503 (2014);
%Marianne Bauer, Meera M. Parish, Tilman Enss,
%Universal equation of state and pseudogap in the two-dimensional Fermi gas,
%Phys. Rev. Lett. 112, 135302 (2014);
%Andrea M. Fischer, Meera M. Parish
%F. Marsiglio, P. Pieri, A. Perali, F. Palestini, G. C. Strinati,
%Pairing effects in the normal phase of a two-dimensional Fermi gas,
%Physical Review B 91, 054509 (2015);

\bibitem{imbalanced}
K. B. Gubbels, and H. T. C. Stoof, 
Imbalanced Fermi gases at unitarity,
Physics Reports {\bf 525}, 255 (2013).



\bibitem{chevyreview}
F. Chevy and C. Mora, Ultra-Cold Polarized Fermi Gases, Rep. Prog. Phys. {\bf 73}, 112401 (2010).


\bibitem{spinliquid} 
Leon Balents, Spin liquids in frustrated magnets,
Nature {\bf 464}, 199 (2010);
Yi Zhou, Kazushi Kanoda, Tai-Kai Ng,
Quantum Spin Liquid States, arXiv:1607.03228;
Lucile Savary and Leon Balents,
Quantum Spin Liquids, arXiv:1601.03742.
%P. Fazekas and P. W. Anderson, 
%On the ground state properties of the anisotropic triangular antiferromagnet
%Philos. Mag. 30, 423 (1974). 

\bibitem{randeria}
M. Randeria in Ultracold Fermi Gases, Proceedings of the International School of Physics Enrico Fermi, Course CLXIV, Varenna, 2006, edited by M. Inguscio, W. Ketterle, and C. Salomon (IOS Press, Amsterdam, 2008).

\bibitem{levin}
K. Levin and Qijin Chen in Ultracold Fermi Gases, Proceedings of the International School of Physics Enrico Fermi, Course CLXIV, Varenna, 2006, edited by M. Inguscio, W. Ketterle, and C. Salomon (IOS Press, Amsterdam, 2008).


\bibitem{lee}
P. A. Lee, N. Nagaosa, and X.-G. Wen, Doping a Mott insulator: Physics of high-temperature superconductivity, Rev. Mod. Phys. {\bf 78}, 17 (2006).  
%A review of high temperature superconductivity from a strongly correlated electron point of view.


\bibitem{leereview}
Patrick A Lee, From high temperature superconductivity to quantum spin liquid: progress in strong correlation physics, Rep. Prog. Phys. {\bf 71}, 012501 (2008).

\bibitem{timusk}
T. Timusk and B Statt, The pseudogap in high-temperature superconductors: an experimental survey,
Rep. Prog. Phys. {\bf 62}, 61 (1999).  
%A review of the experimental knowledge about pseudogaps in high temperature superconductors.







\bibitem{paraconductivity}
L. G. Aslamazov and A. I. Larkin, 
Vliyanie fluktuatsii na svoistva sverkhprovodnika pri temperaturakh vyshe kriticheskoi,
Fiz. Tverd. Tela (Leningrad) {\bf 10}, 1104 (1968) [Sov. Phys. Solid State, Effect of Fluctuations on the Properties of a Superconductor Above the Critical Temperature, {\bf 10}, 875 (1968)];
L. G. Aslamazov and A. A. Varlamov, 
Fluctuation conductivity in intercalated superconductors,
J. Low Temp. Phys. {\bf 38}, 223 (1980).
%S. Hikami and A. I. Larkin, 
%Magnetoresistance Of High Temperature Superconductors,
%Mod. Phys. Lett. B 2, 693 (1988);
%P. P. Freitas et al., Phys. Rev. B 36, 833 (1987);
%T. A. Friedmann, Phys. Rev. B 39, 4258 (1989);
%G. Balestrino, Phys. Rev. B 39, 12 264 (1989); ibid. 46, 14 919 (1992);
%N. Mori, Phys. Rev. B 45, 10 633 (1992); P. Mandal, Physica C 169, 43 (1990).


\bibitem{multi}
W. Eerenstein, N. D. Mathur and J. F. Scott, 
Multiferroic and magnetoelectric materials,
Nature {\bf 442}, 759 (2006).




\bibitem{sachdev} Subir Sachdev, {\em Quantum Phase Transitions}, (Cambridge University Press, 2011).  


\bibitem{giamarchi}
Thierry Giamarchi, {\em Quantum Physics in One Dimension}, (Clarendon Press, Oxford, 2004).


\bibitem{jastrow} R. Jastrow, Many-Body Problem with Strong Forces, Phys. Rev. {\bf 98}, 1479 (1955).  

\bibitem{gutzwiller} Martin C Gutzwiller,  Effect of correlation on the ferromagnetism of transition metals, Phys. Rev. Lett. {\bf 10}, 159 (1963);
Martin C Gutzwiller,  Correlation of Electrons in a Narrow  Band. Phys. Rev. A {\bf 137}, 1726  (1965).


\bibitem{cyrus}
J. Toulouse, C. J. Umrigar,
Optimization of quantum Monte Carlo wave functions by energy minimization
J. Chem. Phys. {\bf 126}, 084102 (2007).

\bibitem{mc}
Jindich Koloren and Lubos Mitas,  
Applications of quantum Monte Carlo methods in condensed systems,
Rep. Prog. Phys. {\bf 74} 026502 (2011).




\bibitem{degennes} P. G. De Gennes, {\em Superconductivity Of Metals And Alloys} (Westview Press, 1999).


\bibitem{Tsuchiya}
 S. Tsuchiya, R. Watanabe, and Y. Ohashi, 
 Pseudogap in Fermionic Density of States in the BCS-BEC Crossover of Atomic Fermi Gases,
 J. Low Temp. Phys. {\bf 158}, 29 (2010);
 R. Watanabe, S. Tsuchiya, and Y. Ohashi, 
 Superfluid density of states and pseudogap phenomenon in the BCS-BEC crossover regime of a superfluid Fermi gas
 Phys. Rev. A {\bf 82}, 043630 (2010);
 S. Tsuchiya, R. Watanabe, and Y. Ohashi, 
 Single-particle properties and pseudogap effects in the BCS-BEC crossover regime of an ultracold Fermi gas above Tc,
 Phys. Rev. A {\bf 80}, 033613 (2009).

\bibitem{ohashispin}
Hiroyuki Tajima, Takashi Kashimura, Ryo Hanai, Ryota Watanabe, and Yoji Ohashi,
Uniform spin susceptibility and spin-gap phenomenon in the BCS-BEC-crossover regime of an ultracold Fermi gas,
Phys. Rev. A {\bf 89}, 033617 (2014);
Hiroyuki Tajima, Ryo Hanai, and Yoji Ohashi,
Strong-coupling corrections to spin susceptibility in the BCS-BEC-crossover regime of a superfluid Fermi gas,
Phys. Rev. A {\bf 93}, 013610 (2016).





\bibitem{magierskigap}
Piotr Magierski, Gabriel Wlazlowski, Aurel Bulgac, and Joaqu\'in E. Drut,
Finite-Temperature Pairing Gap of a Unitary Fermi Gas by Quantum Monte Carlo Calculations,
Phys. Rev. Lett. {\bf 103}, 210403 (2009).





\bibitem{butch}
N. P. Butch, M, C de Andrade, and M. B. Maple,
Resource Letter Scy-3: Superconductivity, Am. J. Phys. {\bf 76}, 106 (2008). 


\bibitem{ag} Abrikosov and Gorkov,
On the Problem of the Knight Shift in Superconductors,
Soviet Phys. JETP {\bf 12}, 1243 (1961).

\bibitem{woolf} Michael A Woolf and F. Reif, Effect of Magnetic Impurities on the Density of States of Superconductors, Phys. Rev.  {\bf 137}, A557 (1965).
%\bibitem{scalapino} S. B. Kaplan, C. C. Chi, D. N. Langenberg, J. J. Chang, S. Jafarey, and D. J. Scalapino, Phys. Rev. B 14, {\bf 4854} (1976)



\bibitem{pe}
A. Damascelli, Z. Hussain, and Zhi-Xun Shen, Rev. Mod. Phys. {\bf 75}, 473 (2003).

\bibitem{tn}
C. Renner, B. Revaz, J.-Y. Genoud, K. Kadowaki, and O. Fischer, Phys. Rev. Lett. {\bf 80}, 149 (1998).

\bibitem{nmr}
Russel E. Walstedt,
{\em The NMR Probe of High-Tc Materials,}
 (Springer, Berlin, 2008).



\bibitem{sh} 
  J. W. Loram, K. Mirza, J. Cooper, W. Liang, and J. Wade, Electronic specific heat of YBa2Cu3O6+x from 1.8 to 300 K, J. Supercondutivity, {\bf 7}, 243 (1994);
G. V. M. Williams, E. M. Haines, and J. L. Tallon, Pair breaking in the presence of a normal-state pseudogap in high-Tc cuprates, Phys. Rev. B, {\bf 57}, 146 (1998).

\bibitem{opticalcond} 
A. V. Puchkov, D. N. Basov, and T. Timusk, The pseudogap state in high-Tc superconductors: An infrared study, J. Phys.: Condens. Matter, {\bf 8}, 10049 (1996);
D. N. Basov, R. Liang, B. Dabrowski, D. A. Bonn, W. N. Hardy, and T. Timusk, Pseudogap and charge dynamics in CuO2 planes in YBCO, Phys. Rev. Lett., {\bf 77}, 4090 (1996)
D. Basov, C. Homes, E. Singley, M. Strongin, T. Timusk,G. Blumberg, and D. van der Marel, Unconventional energetics of the pseudogap state and superconducting state in high-Tc cuprates, Phys. Rev. B {\bf 63}, 134514 (2001).

\bibitem{dccond}
D. Walker, A. P. Mackenzie, and J. R. Cooper, 
Transport properties of zinc-doped YBa$_2$Cu$_3$O$_7$ thin films, 
Phys. Rev. B {\bf 51}, 15653(R) (1995);
 T. Graf, J. M. Lawrene, M. F. Hundley, J. D. Thompson, A. Lacerda, E. Haanappel, M. S. Torikahili, Z. Fisk, and P. C. Canfield, Resistivity, magnetization, and specific heat of YbAgCu$_4$ in high magnetic fields, 
 Phys. Rev. B, {\bf 51}, 15053 (1995); 
 Y. F. Yan, P. Matl, J. M. Harris, and N. P. Ong, Negative magnetoresistance in the c-axis resistivity of Bi$_2$Sr$_2$CaCu$_2$O$_8$ and YBa$_2$Cu$_3$O$_{6+x}$, Phys. Rev. B, {\bf 52}, 751(R) (1995).

\bibitem{susc}
S H Naqib and R S Islam,
Extraction of the pseudogap energy scale from the static magnetic susceptibility of single and double CuO2 plane high-Tc cuprates,
Superconductor Science and Technology, {\bf 21}, 105017 (2008);
R. S. Islam, J. R. Cooper, J. W. Loram, and S. H. Naqib,
Pseudogap and doping-dependent magnetic properties of La$_{2?x}$Sr$_x$Cu$_{1?y}$Zn$_y$O$_4$,
Phys. Rev. B {\bf 81}, 054511 (2010).


\bibitem{nernst}
Y. Wang, L. Li, and N. P. Ong, Phys. Rev. B {\bf 73}, 023510 (2006).

\bibitem{tannernst}
S. Tan and K. Levin,
Nernst Effect and Anomalous Transport in Cuprates: A Preformed-Pair Alternative to the Vortex Scenario,
Phys. Rev. B {\bf 69}, 064510 (2004).


\bibitem{hiddenarpes}
I. M. Vishik, W. S. Lee, R.-H. He, M. Hashimoto, Z. Hussain, T. P. Devereaux, and Z.-X. Shen,
New J. Phys. {\bf 12}, 105008 (2010).

\bibitem{hidden}
Sudip Chakravarty, R. B. Laughlin, Dirk K. Morr, and Chetan Nayak,
Hidden order in the cuprates,
Phys. Rev. B {\bf 63}, 094503 (2001).


\bibitem{lawler}
%Andrej Mesaros, Kazuhiro Fujita, Stephen D. Edkins, Mohammad H. Hamidian, Hiroshi Eisaki, Shin-ichi Uchida,
%J. C. SŽamus Davis, Michael J. Lawler, and Eun-Ah Kim,
%Commensurate 4a0-period charge density modulations throughout the Bi2Sr2CaCu2O8+x pseudogap regime,
%Proc. Nat'l Acad. Sci. 113, 12661 (2016);
	A. R. Schmidt, K. Fujita, E.-A. Kim, M. J. Lawler, H. Eisaki, S. Uchida, D.-H. Lee, J. C. Davis,
Electronic Structure of the Cuprate Superconducting and Pseudogap Phases from Spectroscopic Imaging STM ,
	New Journal of Physics {\bf 13}, 065014 (2011);
	M. J. Lawler, K. Fujita, Jhinhwan Lee, A. R. Schmidt, Y. Kohsaka, Chung Koo Kim, H. Eisaki, S. Uchida, J. C. Davis, J. P. Sethna 
	and Eun-Ah Kim, Intra-unit-cell electronic nematicity of the High-Tc copper-oxide pseudogap states,
	Nature {\bf 466}, 374 (2010).

\bibitem{friendfoe}
M. R. Norman, D. Pines, C. Kallin,
The pseudogap: friend or foe of high Tc?
	Adv. Phys. {\bf 54}, 715 (2005).


\bibitem{quantcrit}
Louis Taillefer,
Superconductivity and Quantum Criticality,
La Physique au Canada, {\bf 67}, 109 (2011)


\bibitem{intertwined}
Eduardo Fradkin, 
Steven A. Kivelson,
and 
John M. Tranquada,
Colloquium: Theory of intertwined orders in high temperature
superconductors,
Rev. Mod. Phys. {\bf 87}, 457 (2015).

\bibitem{nikolic}
Predrag Nikolic and Zlatko Tesanovic, 
Cooper pair insulators and theory of correlated superconductors
Phys. Rev. B {\bf 83}, 064501 (2011).

\bibitem{peierls} R. E. Peierls, {\em Quantum Theory of Solids} (Oxford
Univ. Press, Oxford, England, 1955), p. 108.

\bibitem{mermin}
N. D. Mermin, and H.  Wagner, Absence of Ferromagnetism or Antiferromagnetism in One- or Two-Dimensional Isotropic Heisenberg Models, Phys. Rev. Lett. {\bf 17}, 1133 (1966)

\bibitem{lra}P. A. Lee, T. M. Rice, and P. W. Anderson, Fluctuation Effects at a Peierls Transition, Phys. Rev. Lett. {\bf 31}, 462 (1973).


\bibitem{papatsvelik}
Emiliano Papa, and Alexei M. Tsvelik,
A Pseudogap in the Single-Particle Density of States of a Tomonaga-Luttinger Liquid,
cond-mat/0004007 (2000).

\bibitem{ar}
Elhu Abrahams, Martha Redi, and James W. F. Woo, Effects of Fluctuations on Electronic Properties above the Superconducting Transition, Phys. Rev. B {\bf 1}, 208 (1970);


\bibitem{stajic}
Jelena Stajic, J. N. Milstein, Qijin Chen, M. L. Chiofalo, M. J. Holland, and K. Levin, Nature of superfluidity in ultracold Fermi gases near Feshbach resonances, 
Phys. Rev. A {\bf 69}, 063610 (2004).

\bibitem{landauquantum}
L. D. Landau, and L. M. Lifshitz,
{\em Quantum Mechanics (Non-relativistic Theory),}
(Butterworth-Heinemann, Oxford, 1981).

\bibitem{tiesinga}
E. Tiesinga, B. J. Verhaar, and H. T. C. Stoof, Threshold and resonance phenomena in ultracold ground-state collisions, Phys. Rev. A {\bf 47}, 4114 (1993).


\bibitem{bcs}
J. Bardeen, L. N. Cooper, J. R. Schrieffer, Microscopic Theory of Superconductivity, Phys. Rev. {\bf 106}, 162 (1957);
J. Bardeen, L. N. Cooper, and J. R. Schrieffer, Theory of Superconductivity, Phys. Rev. {\bf 108}, 1175 (1957).


\bibitem{BEC}
 Isaac M. Khalatnikov,
 {\em An Introduction to the Theory of Superfluidity,}
( Westview Press, 2000)

 \bibitem{fermiliquid}
Gordon Baym and Chris Pethick,
{\em Landau Fermi-Liquid Theory: Concepts and Applications,}
(Wiley, 2008).

 


\bibitem{chien} Chih-Chun Chien, Hao Guo, Yan He, K. Levin, Comparative Study of BCS-BEC Crossover Theories above $T_c$: the Nature of the Pseudogap in Ultra-Cold Atomic Fermi Gases,  Phys. Rev. A {\bf 81}, 023622 (2010).

\bibitem{levincompare} K. Levin, Qijin Chen, Chih-Chun Chien, and Yan He, Comparison of Different Pairing Fluctuation Approaches to BCS-BEC Crossover
Ann. Phys. {\bf 325}, 233 (2010).

\bibitem{hucompare}
Hui Hu, Xia-Ji Liu, and Peter D. Drummond,
Comparative study of strong-coupling theories of a trapped Fermi gas at unitarity,
Phys. Rev. A {\bf 77}, 061605(R) (2008);
Hui Hu, Xia-Ji Liu, and Peter D Drummond,
Universal thermodynamics of a strongly interacting Fermi gas: theory versus experiment,
New J. Phys. {\bf 12} , 063038 (2010).





\bibitem{onset}
P. Magierski, G. Wlazlowski, and A. Bulgac, 
Onset of a pseudogap regime in ultracold Fermi gases
Phys. Rev. Lett. {\bf 107}, 145304 (2011).


\bibitem{virialspectraltrap}
H. Hu, X.-J. Liu, P.D. Drummond, H. Dong, 
 Pseudogap Pairing in Ultracold Fermi Atoms,
 Phys. Rev. Lett. {\bf 104}, 240407, (2010) 


\bibitem{strinatireview}
G. Strinati, Pairing fluctuations approach to the BCSÐBEC crossover, in W Zwerger, {\em The BCSÐBEC Crossover and the Unitary Fermi Gas}  (Springer, Heidelberg, 2012) pp 99

\bibitem{levinrfreview}
Qijin Chen, Yan He, Chih-Chun Chien, and K Levin,
Theory of radio frequency spectroscopy experiments in ultracold Fermi gases and their relation to photoemission in the cuprates, Rep. Prog. Phys. {\bf 72}, 122501 (2009).

\bibitem{chinjulienne}
Cheng Chin and Paul S. Julienne,
Radio-frequency transitions on weakly bound ultracold molecules,
Phys. Rev. A {\bf 71}, 012713 (2005).

\bibitem{photoinjection}
Lawrence W. Cheuk, Ariel T. Sommer, Zoran Hadzibabic, Tarik Yefsah, Waseem S. Bakr, and Martin W. Zwierlein,
Spin-Injection Spectroscopy of a Spin-Orbit Coupled Fermi Gas,
Phys. Rev. Lett. {\bf 109}, 095302 (2012)

\bibitem{bcsbecrf}
A. Schirotzek, C.-H. Wu, A. Sommer, M. W. Zwierlein, Phys. Rev. Lett {\bf 102}, 230402 (2009);
J. T. Stewart, J. P. Gaebler, D. S. Jin, Nature {\bf 454}, 744 (2008);
A. Schirotzek, Y.-I. Shin, C. H. Schunck, W. Ketterle,
Phys. Rev. Lett {\bf 101}, 140403 (2008);
Y. Shin, C. H. Schunck, A. Schirotzek, W. Ketterle, Phys. Rev. Lett {\bf 99}, 090403 (2007)

\bibitem{schuncknature}
C. H. Schunck, Y.-I. Shin, A. Schirotzek, W. Ketterle, Nature {\bf 454}, 739 (2008);

\bibitem{greinerrf}
Markus Greiner, Cindy A. Regal, Deborah S. Jin, Phys. Rev. Lett. {\bf 94}, 070403 (2005)

\bibitem{chinrf}
C. Chin, M. Bartenstein, A. Altmeyer, S. Riedl, S. Jochim, J. Hecker Denschlag, and R. Grimm.  
Observation of the Pairing Gap in a Strongly Interacting Fermi Gas
Science {\bf 305}, 1128 (2004).

\bibitem{ketterlerf}
S. Gupta, Z. Hadzibabic, M.W. Zwierlein, C.A. Stan, K. Dieckmann, C.H. Schunck, E.G.M. van Kempen, B.J. Verhaar, and W. Ketterle:
RF Spectroscopy of Ultracold Fermions.
Science {\bf 300}, 1723 (2003).

\bibitem{momentumresolved}
Y. Sagi, T. E. Drake, R. Paudel, R. Chapurin, and D. S. Jin, Probing local quantities in a strongly interacting Fermi gas, Journal of Physics: Conference Series, {\bf 467},  012010 (2013);
A. Perali, F. Palestini, P. Pieri, G. C. Strinati, J. T. Stewart, J. P. Gaebler, T. E.  Drake, and D. S. Jin, Evolution of the Normal State of a Strongly Interacting Fermi Gas from a Pseudogap Phase to a Molecular Bose Gas, Phys. Rev. Lett.  {\bf 106}, 060402  (2011);
J. P. Gaebler,	J. T. Stewart,	T. E. Drake,	D. S. Jin,	A. Perali,	P. Pieri	and G. C. Strinati,
Observation of pseudogap behaviour in a strongly interacting Fermi gas, Nature Physics, {\bf 6}, 569 (2010).
D. S. Jin, J. T. Stewart, and J. P. Gaebler, Photoemission Spectroscopy for Ultracold Atoms, in
R. Cote, P. L. Gould, M. Rozman,  and W. W. Smith, 
 {\em Pushing the Frontiers of Atomic Physics: Proceedings of the XXI International Conference on Atomic Physics,}  (World Scientific, 2009) pp. 213Ð219;
J. T. Stewart, J. P. Gaebler, and D. S. Jin, Using photoemission spectroscopy to probe a strongly interacting Fermi gas, Nature, {\bf 454},  744 (2008).






\bibitem{massignan}
P. Massignan, G. M. Bruun, and H. T. C. Stoof,
Twin peaks in rf spectra of Fermi gases at unitarity,
Phys. Rev. A {\bf 77}, 031601(R) (2008)



\bibitem{generic}
Erich J. Mueller, Generic features of the spectrum of trapped polarized fermions, Phys. Rev. A {\bf 78}, 045601 (2008)


\bibitem{kinnunen}
J. Kinnunen, M. Rodr'guez, and P. T\"orm\"a, Science {\bf 305}, 1131 (2004)



\bibitem{regal}
C. A. Regal, C.C. Ticknor, J. L.  Bohn,  and D. S. Jin, Creation of ultracold molecules from a Fermi gas of atoms, Nature, {\bf 424},  47 (2003);

\bibitem{yanhe}
Yan He, Qijin Chen, and K. Levin, 
Radio-frequency spectroscopy and the pairing gap in trapped Fermi gases
Phys. Rev. A {\bf 72},
011602(R)   (2005);
Yah He, Chih-Chun Chien, Qijin Chen, and K. Levin, 
Radio-frequency spectroscopy of trapped Fermi gases with population imbalance
Phys. Rev. A {\bf 77}, 011602(R) (2008).


\bibitem{ohashi}
Y. Ohashi and A. Griffin, 
Single-particle excitations in a trapped gas of Fermi atoms in the BCS-BEC crossover region,
Phys. Rev. A {\bf 72}, 013601  (2005).





\bibitem{schunck}
C. H. Schunck, Y. Shin, A. Schirotzek, M. W. Zwierlein, and
W. Ketterle, Science {\bf 316}, 867  (2007).



\bibitem{mitspatialresolve}
Andr\'e Schirotzek, Yong-il Shin, Christian H. Schunck, and Wolfgang Ketterle,
Determination of the Superfluid Gap in Atomic Fermi Gases by Quasiparticle Spectroscopy,
Phys. Rev. Lett. {\bf 101} 140403 (2008)



\bibitem{realandk}
Yoav Sagi, Tara E. Drake, Rabin Paudel, Roman Chapurin, and Deborah S. Jin,
Breakdown of the Fermi Liquid Description for Strongly Interacting Fermions,
Phys. Rev. Lett. {\bf 114}, 075301 (2015);


\bibitem{localPES}
Yoav Sagi, Tara E Drake, Rabin Paudel, Roman Chapurin and Deborah S Jin,
Probing local quantities in a strongly interacting Fermi gas,
Journal of Physics: Conference Series {\bf 467}, 012010 (2013) 


\bibitem{strinatiRFAL}
Pierbiagio Pieri, Andrea Perali and Giancarlo Calvanese Strinati,
Enhanced paraconductivity-like fluctuations in the radiofrequency spectra of ultracold Fermi atoms,
Nature Physics {\bf 5}, 736 (2009)

\bibitem{final}
He, Y., Chien, C.-C., Chen, Q. and Levin, K. Temperature and final state effects in radio frequency spectroscopy experiments on atomic Fermi gases. Phys. Rev. Lett. {\bf 102}, 020402 (2009);
Perali, A., Pieri, P. and Strinati, G. C. Competition between final-state and pairing-gap effects in the radio-frequency spectra of ultracold Fermi atoms. Phys. Rev. Lett. {\bf 100}, 010402 (2008);
S. Basu and E. J. Mueller,  Final-state effects in the radio frequency spectrum of strongly interacting fermions. Phys. Rev. Lett. {\bf 101}, 060405 (2008);
Z. Yu and G. Baym, 
Spin-correlation functions in ultracold paired atomic-fermion systems: Sum rules, self-consistent approximations, and mean fields,
Phys. Rev. A {\bf 73}, 063601 (2006);
P. Pieri, A. Perali, G. C. Strinati, S. Riedl, M. J. Wright, A. Altmeyer, C. Kohstall, E. R. S‡nchez Guajardo, J. Hecker Denschlag, and R. Grimm,
Pairing-gap, pseudogap, and no-gap phases in the radio-frequency spectra of a trapped unitary $^6$Li gas,
Phys. Rev. A {\bf 84}, 011608(R) (2011);
M. Punk and W. Zwerger,
Theory of rf-Spectroscopy of Strongly Interacting Fermions,
Phys. Rev. Lett. {\bf 99}, 170404 (2007);
Gordon Baym, C. J. Pethick, Zhenhua Yu, and Martin W. Zwierlein,
Coherence and Clock Shifts in Ultracold Fermi Gases with Resonant Interactions,
Phys. Rev. Lett. {\bf 99}, 190407 (2007).



\bibitem{palestinilifetimes}
F. Palestini, A. Perali, P. Pieri, and G. C. Strinati,
Dispersions, weights, and widths of the single-particle spectral function in the normal phase of a Fermi gas,
Phys. Rev. B {\bf 85}, 024517 (2012).


\bibitem{lifetime}
Matthew D. Reichl and Erich J. Mueller,
Quasiparticle dispersions and lifetimes in the normal state of the BCS-BEC crossover,
Phys. Rev. A {\bf 91} 043627 (2015).





\bibitem{hydro}
J. A. Joseph, E. Elliott, J. E. Thomas, 
Shear viscosity of a universal Fermi gas near the superfluid phase transition,
Phys. Rev. Lett. {\bf 115}, 020401 (2015); E. Elliott, J. A. Joseph, J. E.Thomas, 
Anomalous minimum in the shear viscosity of a Fermi gas, Phys. Rev. Lett. {\bf 113}, 020406 (2014).


\bibitem{parisEOS}
S. Nascimbne, N. Navon, K. J. Jiang, F. Chevy and C. Salomon,
Exploring the thermodynamics of a universal Fermi gas,
Nature {\bf 463}, 1057 (2010).

\bibitem{MITEOS}
K. Van Houcke,	F. Werner,	E. Kozik,	N. ProkofÕev,	B. Svistunov,	M. J. H. Ku,	A. T. Sommer,	L. W. Cheuk,	A. Schirotzek	 and M. W. Zwierlein,
Feynman diagrams versus Fermi-gas Feynman emulator,
Nature Physics {\bf 8}, 366 (2012).



\bibitem{levinEOS}
Yan He, Chih-Chun Chien, Qijin Chen, K. Levin,
Thermodynamics and superfluid density in BCS-BEC crossover with and without population imbalance,
Phys. Rev. B. {\bf 76}, 224516 (2007);
Qijin Chen, Jelena Stajic, K. Levin,
Thermodynamics of Interacting Fermions in Atomic Traps,
Phys. Rev. Lett. {\bf 95}, 260405 (2005); 
Jelena Stajic, Qijin Chen, Kathryn Levin,
Density Profiles of Strongly Interacting Trapped Fermi Gases,
Phys. Rev. Lett. {\bf 94}, 060401 (2005).



 \bibitem{chienlevin}
Chih-Chun Chien, and K. Levin, 
Fermi-liquid theory of ultracold trapped Fermi gases: Implications for pseudogap physics and other strongly correlated phases,
Phys. Rev. A {\bf 82}, 013603 (2010).  



\bibitem{levinFL}
Chih-Chun Chien, K. Levin, 
Fermi liquid theory of ultra-cold trapped Fermi gases: Implications for Pseudogap Physics and Other Strongly Correlated Phases,
Phys. Rev. A {\bf 82}, 013603 (2010)



\bibitem{ketterleeos}
Y. Shin,
Determination of the equation of state of a polarized Fermi gas at unitarity. 
Phys. Rev. A {\bf 77}, 041603(R) (2008).


\bibitem{strinatiEOS}
A. Perali, P. Pieri, L. Pisani, G.C. Strinati,
BCS-BEC crossover at finite temperature for superfluid trapped Fermi atoms,
Phys. Rev. Lett. {\bf 92}, 220404 (2004).



\bibitem{zwierleinhomo}
Biswaroop Mukherjee, Zhenjie Yan, Parth B. Patel, Zoran Hadzibabic, Tarik Yefsah, Julian Struck, 
and Martin W. Zwierlein,
Homogeneous Atomic Fermi Gases, arXiv:1610.10100 (2016).




\bibitem{thermo}
Joseph Kinast, Andrey Turlapov, John E. Thomas, Qijin Chen, Jelena Stajic, and Kathryn Levin,
Heat Capacity of a Strongly-Interacting Fermi Gas,
Science {\bf 307}, 1296 (2005);
Le Luo and J. E. Thomas,
Thermodynamic Measurements in a Strongly Interacting Fermi Gas,
J Low Temp Phys. {\bf 154}, 1 (2009)


\bibitem{compress}
Y.R. Lee, M.-S. Heo, J.H. Choi, T.T. Wang, C.A. Christensen, T.M. Rvachov, and W. Ketterle: 
Compressibility of an Ultracold Fermi Gas with Repulsive Interactions. 
Phys. Rev. A {\bf 85}, 063615 (2012). 


\bibitem{mitimage}
M.W. Zwierlein, C.H. Schunck, A. Schirotzek, W. Ketterle,
Direct Observation of the Superfluid Phase Transition in Ultracold Fermi Gases. 
Nature {\bf 442}, 54-58 (2006).


\bibitem{criticalvel}
D. E. Miller, J. K. Chin and C. A. Stan, Y. Liu, W. Setiawan, C. Sanner and W. Ketterle: 
Critical velocity for superfluid flow across the BEC-BCS crossover. 
Phys. Rev. Lett. {\bf 99}, 070402 (2007). 





\bibitem{lingham}
M. G. Lingham, K. Fenech, T. Peppler, S. Hoinka, P. Dyke, P. Hannaford, and C. J. Vale,
Bragg spectroscopy of strongly interacting Fermi gases,
Journal of Modern Optics, {\bf 63},  1783 (2016);
M. G. Lingham, K. Fenech, S. Hoinka, and C. J. Vale,
Local Observation of Pair Condensation in a Fermi Gas at Unitarity,
 Phys. Rev. Lett. {\bf 112}, 100404 (2014)
 
 
 


 
\bibitem{bragg}
 G. Veeravalli, E. Kuhnle, P. Dyke, and C. J. Vale, Phys. Rev. Lett. {\bf 101}, 250403 (2008).



\bibitem{bruun}
G. M. Bruun and Gordon Baym.
Bragg spectroscopy of cold atomic Fermi gases.  Phys.  Rev.  A {\bf 74},
033623 (2006).


\bibitem{fluc}
C. Sanner, E.J. Su, A. Keshet, W. Huang, J. Gillen, R. Gommers, and W. Ketterle: 
Speckle Imaging of Spin Fluctuations in a Strongly Interacting Fermi Gas. 
Phys. Rev. Lett. {\bf 106}, 010402 (2011). 





%\bibitem{bcsbecexp}
%L. Luo, B. Clancy, J. Joseph, J. Kinast, J. E. Thomas, Phys. Rev. Lett. {\bf 98}, 080402 (2007);
%G. B. Partridge, K. E. Strecker, R. I. Kamar, M. W. Jack, R. G. Hulet, Phys. Rev. Lett. {\bf 95}, 020404 (2005);
%T. Bourdel, L. Khaykovich, J. Cubizolles, J. Zhang, F. Chevy, M. Teichmann, L. Tarruell, S. J. J. M. F. Kokkelmans, and C. Salomon Phys. Rev. Lett. {\bf 93}, 050401 (2004).  These are the seminal studies of the BCS-BEC crossover


%\bibitem{critpol}
%Y.-I. Shin, C. H. Schunck, A. Schirotzek, and W.  Ketterle,
%Nature {\bf 451}, 689 (2008).
%%C. Lobo, A. Recati, S. Giorgini, and  S. Stringari, Phys. Rev. Lett. {\bf 97}, 200403 (2006).  
%Critical polarization.
%\bibitem{zw}
%Andre Schirotzek, Cheng-Hsun Wu, Ariel Sommer, Martin W. Zwierlein,
%Physical Review Letters 102, 230402 (2009).
%%\bibitem{photoasssociation}
%%M. Junker, D. Dries, C. Welford, J. Hitchcock, Y. P. Chen, and R. G. Hulet, "Photoassociation of a Bose-Einstein Condensate near a Feshbach Resonance," Physical Review Letters 101, 060406 (2008).

%
%\bibitem{trivedi}
%Vamsi K. Akkineni, D.M. Ceperley, and Nandini Trivedi, 
%Pairing and superfluid properties of dilute fermion gases at unitarity
%Phys. Rev. B {\bf 76}, 165116 (2006). 




\bibitem{trivediranderia}
Trivedi and Randeria, 
Deviations from Fermi-Liquid Behavior above Tc in 2D Short Coherence Length Superconductors,
Phys. Rev. Lett. {\bf 75}, 312 (1995)

\bibitem{exptsusc}
S. Nascimbne, N. Navon, S. Pilati, F. Chevy, S. Giorgini, A. Georges, and C. Salomon,
Fermi-Liquid Behavior of the Normal Phase of a Strongly Interacting Gas of Cold Atoms,
Phys. Rev. Lett. {\bf 106}, 215303 (2011).

\bibitem{mitsusc}
Ariel Sommer, Mark Ku, Giacomo Roati, Martin W. Zwierlein
Universal Spin Transport in a Strongly Interacting Fermi Gas
Nature {\bf 472}, 201 (2011)


\bibitem{mcdip}
Gabriel Wlazlowski, Piotr Magierski, Joaqu\'in E. Drut, Aurel Bulgac, and Kenneth J. Roche,
Cooper Pairing Above the Critical Temperature in a Unitary Fermi Gas,
Phys. Rev. Lett. {\bf 110}, 090401 (2013).



\bibitem{dynspin}
S. Hoinka, M. Lingham, M. Delehaye, and C. J. Vale,
Dynamic Spin Response of a Strongly Interacting Fermi Gas,
Phys. Rev. Lett. {\bf 109}, 050403 (2012).

\bibitem{dynspintheory}
%Y. Nishida, Phys. Rev. A 85, 053643 (2012);
 H. Hu and X.-J. Liu, Universal dynamic structure factor of a strongly correlated Fermi gas, Phys. Rev. A {\bf 85}, 023612 (2012).
S. Stringari, Density and Spin Response Function of a Normal Fermi Gas at Unitarity, Phys. Rev. Lett. {\bf 102}, 110406 (2009).
 G. M. Bruun and G. Baym, Bragg spectroscopy of cold atomic Fermi gases, Phys. Rev. A {\bf 74}, 033623 (2006).
 H. Guo, C.-C. Chien, and K. Levin, Establishing the Presence of Coherence in Atomic Fermi Superfluids: Spin-Flip and Spin-Preserving Bragg Scattering at Finite Temperatures, Phys. Rev. Lett. {\bf 105},
120401 (2010).

\bibitem{dynspin2}
R. Combescot, S. Giorgini, and S. Stringari,
Molecular signatures in the structure factor of an interacting Fermi gas,
Europhys. Lett. {\bf 75}, 695 (2006).


\bibitem{transportreview}
Allan Adams, Lincoln D. Carr, Thomas Schaefer, Peter Steinberg and John E. Thomas,
Strongly Correlated Quantum Fluids: Ultracold Quantum Gases, Quantum Chromodynamic Plasmas, and Holographic Duality,
New J. Phys. {\bf 14}, 115009 (2012).



\bibitem{anisoexp}
K. M. OÕHara, S. L. Hemmer, M. E. Gehm, S. R. Granade, J. E. Thomas,
Observation of a Strongly
Interacting Degenerate Fermi
Gas of Atoms,
Science {\bf 298}, 2179 (2002).

\bibitem{colmodes}
J. Kinast, A. Turlapov, and J. E. Thomas, 
Breakdown of hydrodynamics in the radial breathing mode of a strongly interacting Fermi gas,
Phys. Rev. A {\bf 70}, 051401(R) (2004);
J. Kinast, S. L. Hemmer, M. E. Gehm, A. Turlapov, and J. E. Thomas, 
Evidence for Superfluidity in a Resonantly Interacting Fermi Gas, 
Phys. Rev. Lett. {\bf 92}, 150402  (2004);
  M. Bartenstein, A. Altmeyer, S. Riedl, S. Jochim, C. Chin, J. H. Denschlag, and R. Grimm, 
Collective Excitations of a Degenerate Gas at the BEC-BCS Crossover, Phys. Rev. Lett. {\bf 92}, 203201  (2004);
  J. Kinast, A. Turlapov, and J. E. Thomas, Damping of a Unitary Fermi Gas, Phys. Rev. Lett. {\bf 94}, 170404  (2005).



\bibitem{expansion}
E. Elliott, J. A. Joseph, and J. E. Thomas, Observation of Conformal Symmetry Breaking and Scale Invariance in Expanding Fermi Gases,
Phys. Rev. Lett. {\bf 112}, 040405 (2014)

\bibitem{universal}
C. Cao, E. Elliott, J. Joseph, H. Wu, J. Petricka, T. Sch\"afer, and J. E. Thomas, 
Universal Quantum Viscosity in a Unitary Fermi Gas,
Science, {\bf 331}, 58 (2011).




\bibitem{damping}
C Cao, E Elliott, H Wu and J E Thomas,
Searching for perfect fluids: quantum viscosity in a universal Fermi gas,
New J. Phys. {\bf 13}, 075007 (2011);
E. Elliott, J.?A. Joseph, and J. E. Thomas
Anomalous Minimum in the Shear Viscosity of a Fermi Gas,
Phys. Rev. Lett. {\bf 113}, 020406 (2014)



\bibitem{thywissenspin}
S. Trotzky, S. Beattie, C. Luciuk, S. Smale, A. B. Bardon, T. Enss, E. Taylor, Shizhong Zhang, J. H. Thywissen
Observation of the Leggett-Rice effect in a unitary Fermi gas
Physical Review Letters, {\bf 114}, 015301 (2015);
A. B. Bardon, S. Beattie, C. Luciuk, W. Cairncross, D. Fine, N. S. Cheng, G. J. A. Edge, E. Taylor, Shizhong Zhang, S. Trotzky, J. H. Thywissen
Transverse Demagnetization Dynamics of a Unitary Fermi Gas
Science {\bf 344}, 722 (2014);

\bibitem{zwierleinspin}
Ariel Sommer, Mark Ku, and Martin W. Zwierlein
Spin Transport in Polaronic and Superfluid Fermi Gases
New J. Phys. {\bf 13}, 055009 (2011);
%Ariel Sommer, Mark Ku, Giacomo Roati, and Martin W. Zwierlein
%Universal Spin Transport in a Strongly Interacting Fermi Gas
%Nature 472, 201-204 (2011);


\bibitem{othertransport}
A. Sommer, M. Ku, G. Roati, and M. W. Zwierlein, 
Universal spin transport in a strongly interacting Fermi gas
Nature 472, 201 (2011).


\bibitem{spincurrents}
X. Du, Y. Zhang, J. Petricka, and J. E. Thomas,
Controlling Spin Current in a Trapped Fermi Gas,
Phys. Rev. Lett. {\bf 103}, 010401 (2009);
X. Du, L. Luo, B. Clancy, and J. E. Thomas,
Observation of Anomalous Spin Segregation in a Trapped Fermi Gas,
Phys. Rev. Lett. {\bf 101}, 150401 (2008).


\bibitem{levinspin}
Dan Wulin, Hao Guo, Chih-Chun Chien, Kathryn Levin,
Spin Transport in Cold Fermi gases: A Pseudogap Interpretation of Spin Diffusion Experiments at Unitarity,
Phys. Rev. A {\bf 83}, 061601(R) (2011)


%\bibitem{collectivemodetheory}
%Theja N. De Silva and Erich J. Mueller,
%Collective oscillations of a Fermi gas near a Feshbach resonance,
%Phys. Rev. A 72, 063614 (2005);
%S. Stringari, Europhys. Lett. 65, 749  2004 .
%  H. Hu, A. Minguzzi, X. J. Liu, and M. P. Tosi, Phys. Rev. Lett.
%93, 190403  2004 .
%  H. Heiselberg, Phys. Rev. Lett. 93, 040402  2004 .
%  Y. E. Kim and A. L. Zubarev, Phys. Rev. A 70, 033612  2004 .
%   N. Manini and L. Salasnich, Phys. Rev. A 71, 033625  2005 .
%  G. E. Astrakharchik, R. Combescot, X. Leyronas, and S. Strin-
%gari, Phys. Rev. Lett. 95, 030404  2005 .
%   A. Bulgac and G. F. Bertsch, Phys. Rev. Lett. 94, 070401
% 2005 .
%   Y. Ohashi and A. Griffin, eprint cond-mat/0503641.


\bibitem{junction}
S. Krinner, M. Lebrat, D. Husmann, C. Grenier, J. - P. Brantut, and T. Esslinger:
Mapping out spin and particle conductances in a quantum point contact
Proc. Natl. Acad. Sci. U.S.A.
%Proceedings of the National Academy of Sciences 
{\bf 113}, 8144 (2016);
D. Husmann, S. Uchino, S. Krinner, M. Lebrat, T. Giamarchi, T. Esslinger, and J. - P. Brantut:
Connecting strongly correlated superfluids by a quantum point contact,
Science {\bf 350}, 1498 (2015);
S. Krinner, D. Stadler, D. Husmann, J. - P. Brantut, and T. Esslinger:
Observation of quantized conductance in neutral matter
Nature {\bf 517}, 64 (2015);
D. Stadler, S. Krinner, J. Meineke, J. - P. Brantut, and T. Esslinger:
Observing the Drop of Resistance in the Flow of a Superfluid Fermi Gas
Nature {\bf 491}, 736 (2012);
J. - P. Brantut, J. Meineke, D. Stadler, S. Krinner, and T. Esslinger:
Conduction of Ultracold Fermions Through a Mesoscopic Channel
Science {\bf 337}, 1069 (2012).



\bibitem{demlerpair}
M\'arton Kan\'asz-Nagy, Leonid Glazman, Tilman Esslinger, Eugene A. Demler,
Anomalous conductances in an ultracold quantum wire,  arXiv:1607.02509.

\bibitem{uedapair}
Shun Uchino, Masahito Ueda, 
Anomalous Transport in a Superfluid Fluctuation Regime,
arXiv:1608.01070.

\bibitem{zhangpair}
Boyang Liu, Hui Zhai, Shizhong Zhang,
Anomalous Conductance of a Unitary Fermi Gas through a Quantum Point Contact, arXiv:1608.05909.



\bibitem{tan}
S. Tan, Energetics of a strongly correlated Fermi gas, Ann. Phys. {\bf 323}, 2952 (2008); 
Large momentum part of a strongly correlated Fermi gas
2971 (2008); 
S. Tan, 
Generalized virial theorem and pressure relation for a strongly correlated Fermi gas, 
Ann. Phys. {\bf 323},
2987 (2008).

\bibitem{othercontact}
E. Braaten, Universal relations for fermions with large
scattering lengths, in W Zwerger, {\em The BCS-BEC Crossover and the Unitary Fermi Gas}  (Springer, Heidelberg, 2012)  pp 193Ð231;
E Braaten and L. Platter, 
Exact Relations for a Strongly Interacting Fermi Gas from the Operator Product Expansion
Phys. Rev. Lett. {\bf 100}, 205301 (2008);
Eric Braaten, Daekyoung Kang, and Lucas Platter,
Short-Time Operator Product Expansion for rf Spectroscopy of a Strongly Interacting Fermi Gas,
Phys. Rev. Lett. {\bf 104}, 223004 (2010);
F. Werner, L. Tarruel, and Y. Castin, Number of closed-channel molecules in the BEC-BCS crossover, Eur. Phys. J B {\bf 68}, 401 (2009);
S. Zhang and A. J. Leggett, Universal properties of the ultracold Fermi gas, Phys. Rev. A {\bf 79}, 023601 (2009);
Igor Boettcher, Sebastian Diehl, Jan Pawlowski, and Christof Wetterich,
Tan contact and universal high momentum behavior of the fermion propagator in the BCS-BEC crossover,
Phys. Rev. A {\bf 87}, 023606 (2013).



\bibitem{jintof}
T. E. Drake, Y. Sagi, R. Paudel, J. T. Stewart, J. P. Gaebler, D. S. Jin,
Direct observation of the Fermi surface in an ultracold atomic gas,
Phys. Rev. A {\bf 86}, 031601 (2012).



\bibitem{palestinicontact}
F. Palestini, A. Perali, P. Pieri, and G. C. Strinati, 
Temperature and coupling dependence of the universal contact intensity for an ultracold Fermi gas,
Phys. Rev. A {\bf 82}, 021605 (2010).

\bibitem{goulko}
Olga Goulko and Matthew Wingate, Numerical study of the unitary Fermi gas across the superfluid transition,
Phys. Rev. A {\bf 93}, 053604 (2016).




\bibitem{bdmcontact}
K. Van Houcke, F. Werner, E. Kozik, N. Prokof'ev, B. Svistunov,
Contact and Momentum Distribution of the Unitary Fermi Gas by Bold Diagrammatic Monte Carlo,
arXiv:1303.6245


\bibitem{balcontact}
Z. Yu, G. M. Bruun, and G. Baym, 
Short-range correlations and entropy in ultracold-atom Fermi gases
Phys. Rev. A {\bf 80}, 023615 (2009).

\bibitem{kuhnle}
E. D. Kuhnle, H. Hu, X.-J. Liu, P. Dyke, M. Mark, P. D. Drummond, P. Hannaford, and C. J. Vale, 
Universal Behavior of Pair Correlations in a Strongly Interacting Fermi Gas
Phys. Rev. Lett. {\bf 105}, 070402 (2010);
E. D. Kuhnle, S. Hoinka, P. Dyke, H. Hu, P. Hannaford, and C. J. Vale,
Temperature Dependence of the Universal Contact Parameter in a Unitary Fermi Gas,
PRL {\bf 106}, 170402 (2011).


\bibitem{jincontact}
Yoav Sagi, Tara E. Drake, Rabin Paudel, and Deborah S. Jin,
Measurement of the Homogeneous Contact of a Unitary Fermi Gas,
Phys. Rev. Lett. {\bf 109}, 220402 (2012).


\bibitem{jincontact2}
J. T. Stewart, J. P. Gaebler, T. E. Drake, and D. S. Jin,
Verification of Universal Relations in a Strongly Interacting Fermi Gas
Phys. Rev. Lett {\bf 104}, 235301 (2010).



\bibitem{photoassociationreview}
R.G. Hulet, Photoassociation of Ultracold Atoms, in James F. Babb, Kate Kirby, and Hossein Sadeghpour,
{\em Proceedings of the Dalgarno Celebratory Simposium}
(Imperial College Press, London 2010).


\bibitem{photoassociation}  
Jones, K. M., E. Tiesinga, P. D. Lett, and P. S. Julienne, Photoassociation spectroscopy of ultracold atoms: Long- range molecules and atomic scattering, Rev. Mod. Phys. {\bf 78}, 483 (2006).

\bibitem{thermalphoto}
R. Wester, S. D. Kraft, M. Mudrich, M. U. Staudt, J. Lange, N. Vanhaecke, O. Dulieu, M. Weidemuller,
Photoassociation inside an optical dipole trap: absolute rate coefficients and Franck-Condon factors,
Appl. Phys. B {\bf 79}, 993 (2004);
S. D. Kraft, M. Mudrich, M. U. Staudt, J. Lange, O. Dulieu, R. Wester, and M. Weidem\"uller
 Saturation of Cs$_2$ photoassociation in an optical dipole trap,
 Phys. Rev. A {\bf 71}, 013417 (2005);
 P. G. Mickelson, Y. N. Martinez, A. D. Saenz, S. B. Nagel, Y. C. Chen, T. C. Killian, P. Pellegrini, and R. C\^ot\'e,
 Phys. Rev. Lett. {\bf 95}, 223002 (2005).

\bibitem{bosephoto}
M. Junker, D. Dries, C. Welford, J. Hitchcock, Y. P. Chen, and R. G. Hulet, Photoassociation of a Bose-Einstein Condensate near a Feshbach Resonance, Phys. Rev. Lett. {\bf 101}, 060406 (2008);
C. McKenzie, J. Hecker Denschlag, H. HŠffner, A. Browaeys, Lu's E. E. de Araujo, F. K. Fatemi, K. M. Jones, J. E. Simsarian, D. Cho, A. Simoni, E. Tiesinga, P. S. Julienne, K. Helmerson, P. D. Lett, S. L. Rolston, and W. D. Phillips, 
Photoassociation of Sodium in a Bose-Einstein Condensate
Phys. Rev. Lett. {\bf 88}, 120403 (2002);
Ionut D. Prodan, Marin Pichler, Mark Junker, Randall G. Hulet, and John L. Bohn,
Intensity Dependence of Photoassociation in a Quantum Degenerate Atomic Gas,
Phys. Rev. Lett. {\bf 91}, 080402 (2003).


\bibitem{ultramol}
Lincoln D Carr, David DeMille, Roman V Krems, and Jun Ye,
Cold and ultracold molecules: science, technology and applications,
New J. Phys. {\bf 11}, 055049 (2009)


\bibitem{fermiphoto}
G. B. Partridge, K. E. Strecker, R. I. Kamar, M. W. Jack, and R. G. Hulet,
Phys. Rev. Lett. {\bf 95}, 020404 (2005).


\bibitem{inelastic}
X. Du, Y. Zhang, and J. E. Thomas, 
Inelastic Collisions of a Fermi Gas in the BEC-BCS Crossover, Phys. Rev. Lett. {\bf 102}, 250402 (2009).



\bibitem{sweep}
C. A. Regal, M. Greiner,  and D. S. Jin, Observation of Resonance Condensation of Fermionic Atom Pairs, Phys. Rev. Lett. {\bf 92}, 040403 (2004);
M. Greiner, C.A. Regal, and D. S. Jin, Emergence of a molecular BoseÐEinstein condensate from a Fermi gas, Nature, {\bf 426}, 537 (2003);
M.W. Zwierlein, C.H. Schunck, C.A. Stan, S.M.F. Raupach, and W. Ketterle,
Formation Dynamics of a Fermion Pair Condensate,
Phys. Rev. Lett. {\bf 94}, 180401 (2005). 

\bibitem{vortsweep}
M.W. Zwierlein, J.R. Abo-Shaeer, A. Schirotzek, C.H. Schunck, and W. Ketterle:
Vortices and Superfluidity in a Strongly Interacting Fermi Gas.
Nature {\bf 435}, 1047 (2005).

\bibitem{solitonsweep}
Tarik Yefsah,	Ariel T. Sommer,	Mark J. H. Ku,	Lawrence W. Cheuk,	Wenjie Ji,	Waseem S. Bakr	and Martin W. Zwierlein,
Heavy solitons in a fermionic superfluid,
Nature {\bf 499}, 426 (2013).

\bibitem{sweepmodel}
Roberto B. Diener, Tin-Lun Ho, Projecting Fermion Pair Condensates into Molecular Condensates
arXiv:cond-mat/0404517

\bibitem{sweepmodel2}
A. Perali, P. Pieri, G.C. Strinati,
Extracting the condensate density from projection experiments with Fermi gases,
Phys. Rev. Lett. {\bf 95}, 010407 (2005).


\bibitem{eagles}D. M. Eagles, 
Possible pairing without superconductivity at low carrier concentrations in bulk and thin-film superconducting semiconductors,
Phys. Rev. {\bf 186}, 456 (1969).

\bibitem{leggett}
A. J. Leggett, Diatomic molecules and Cooper pairs.   in {\em Modern Trends in the Theory of Condensed Matter}, ed. A. Pekalski and J. Przystawa (Springer, Berlin, 1980) p 14;


\bibitem{nozieres}
P. Nozieres and S. Schmitt-Rink, Bose condensation in an attractive fermion gas: from weak to strong coupling superconductivity. J. Low Temp. Phys, {\bf 59}, 195 (1985).


\bibitem{earlyranderia}
C. A. R. Sa de Melo, M. Randeria, and J. R. Engelbrecht. Crossover from BCS to Bose superconductivity: Transition temperature and time-dependent Ginzburg-Landau theory. Phys. Rev. Lett., {\bf 71}, (1993);
J. R. Engelbrecht, M. Randeria, and C. A. R. Sa de Melo, BCS to Bose crossover: Broken-symmetry state. Phys. Rev. B, {\bf 55}, 15153 (1997);
M. Randeria. Precursor pairing correlations and pseudogaps, in G. Iadonisi, J. R. Schrieffer, and M. L. Chiafalo,  {\em Proceedings of the International School of Physics ÒEnrico FermiÓ Course CXXXVI on High Temperature Superconductors,} (IOS Press, 1998).


\bibitem{micnas}
Micnas, R., and S. Robaszkiewicz,  Cond. Matt. Phys. {\bf 1}, 89 (1998).
Micnas, R., J. Ranninger, and S. Robaszkiewicz,  Rev. Mod. Phys. {\bf 62}, 113 (1990).

\bibitem{ranninger}
J. Ranninger, S. Robaszkiewicz, A. Sulpice, and R. Tournier,
Are There Heavy Bosons in Heavy-Fermion Systems and $^3$He?,
Europhysics Letters, {\bf 3}, 347 (1987);
Ranninger, J., and J. M. Robin,  Phys. Rev. B {\bf 53}, 11961(R)  (1996).



\bibitem{bertsch}
Bertsch,G.F. The Many-Body Challenge Problem (MBX) (1999).


\bibitem{castin}
Yvan Castin and F\'elix Werner, The Unitary Gas and its Symmetry Properties,  in W Zwerger, {\em The BCSÐBEC Crossover and the Unitary Fermi Gas}  (Springer, Heidelberg, 2012).

\bibitem{trentozeroT}
G. E. Astrakharchik, J. Boronat, J. Casulleras, and S. Giorgini,
Phys. Rev. Lett. {\bf 93}, 200404  (2004).


\bibitem{polaronspectrum}
Olga Goulko, Andrey S. Mishchenko, Nikolay ProkofÕev, and Boris Svistunov,
Dark Continuum in the Spectral Function of the Resonant Fermi Polaron,  arXiv:1603.06963.

\bibitem{carlsonzeroT}
J. Carlson, S.-Y. Chang, V. R. Pandharipande, and K. E. Schmidt, Superfluid Fermi Gases with Large Scattering Length, Phys. Rev. Lett. {\bf 91}, 050401 (2003);
J. Carlson, J. Morales, Jr., V. R. Pandharipande, and D. G. Ravenhall,
Quantum Monte Carlo calculations of neutron matter
Phys. Rev. C 68, 025802 (2003);
S. Y. Chang, V. R. Pandharipande, J. Carlson, and K. E. Schmidt,
Quantum Monte Carlo studies of superfluid Fermi gases,
Phys. Rev. A 70, 043602 (2004);
J. Carlson and Sanjay Reddy,
Asymmetric Two-Component Fermion Systems in Strong Coupling,
Phys. Rev. Lett. 95, 060401
(2005);
Alexandros Gezerlis and J. Carlson,
Strongly paired fermions: Cold atoms and neutron matter,
Phys. Rev. C 77, 032801(R) (2008).

\bibitem{nosign}
J. Carlson, Stefano Gandolfi, Kevin E. Schmidt, and Shiwei Zhang,
Auxiliary-field quantum Monte Carlo method for strongly paired fermions,
Phys. Rev. A {\bf 84}, 061602(R) (2011).



\bibitem{bulgac}
%Aurel Bulgac, Joaqu\,in E. Drut, and Piotr Magierski,
%Spin 1/2 Fermions in the Unitary Regime: A Superfluid of a New Type,
%Phys. Rev. Lett. {\bf 96}, 090404 (2006);
%Aurel Bulgac, Joaqu'n E. Drut, and Piotr Magierski,
%Quantum Monte Carlo simulations of the BCS-BEC crossover at finite temperature,
%Phys. Rev. A {\bf 78}, 023625 (2008);
Aurel Bulgac, Joaquin E. Drut, Piotr Magierski,
Spin 1/2 Fermions in the Unitary Regime at Finite Temperature,
Int. J. Mod. Phys. B {\bf 20}, 5165 (2006);
Aurel Bulgac, Michael McNeil Forbes and Piotr Magierski, The Unitary Fermi Gas: From Monte Carlo to Density Functionals, in
Zwerger, {\em The BCS-BEC Crossover and the Unitary Fermi Gas}, (Springer, Heidelberg,  2011)
%Bulgac,A.,Yu,Y.:Phys.Rev.Lett.91(19),190404(2003)



\bibitem{wlazlowski}
J.E. Drut, T.A. Lahde, and T. Ten, Momentum distribution and contact of the unitary Fermi gas, Phys. Rev. Lett. {\bf 106}, 205302 (2011);
G. Wlazlowski and P. Magierski, Quantum Monte Carlo study of dilute neutron matter at finite temperatures, Phys. Rev. C {\bf 83}, 012801(R) (2011);
A.  Bulgac, J.E. Drut, and P. Magierski, Quantum Monte Carlo simulations of the BCS-BEC crossover at finite temperatures, Phys. Rev. A {\bf 78}, 023625 (2008);
A. Bulgac, J.E. Drut, and P. Magierski, Thermodynamics of a trapped unitary Fermi gas, Phys. Rev. Lett. {\bf 99}, 120401 (2007); 
A. Bulgac, J.E. Drut, and P. Magierski, Spin 1/2 fermions in the unitary regime: a superfluid of a new type, Phys. Rev. Lett. {\bf 96}, 090404 (2006).


\bibitem{wingate}
M. Wingate, Critical temperature for fermion pairing using lattice field theory, cond-mat/0502372 (2005).

\bibitem{leeschaefer}
D. Lee and T. Schaefer, Cold dilute neutron matter on the lattice. II. Results in the unitary limit, Phys. Rev. C 73, 015202 (2006). 



\bibitem{determinantmc}
Evgeni Burovski, Nikolay ProkofÕev, Boris Svistunov, and Matthias Troyer,
Critical Temperature and Thermodynamics of Attractive Fermions at Unitarity,
Phys. Rev. Lett. 96, 160402 (2006);
Evgeni Burovski, Evgeny Kozik, Nikolay ProkofÕev, Boris Svistunov, and Matthias Troyer,
Critical Temperature Curve in BEC-BCS Crossover,
Phys. Rev. Lett. {\bf 101}, 090402 (2008).

\bibitem{trivedi}
Vamsi K. Akkineni, D.M. Ceperley, and Nandini Trivedi, 
Monte Carlo calculation of transition temperature and formation of pairing correlations,
Phys. Rev. B {\bf 76}, 165116 (2006). 

\bibitem{amherstMIT}
K. Van Houcke,	F. Werner,	E. Kozik,	N. ProkofÕev,	B. Svistunov,	M. J. H. Ku,	A. T. Sommer,	L. W. Cheuk,	A. Schirotzek	
and M. W. Zwierlein, Feynman diagrams versus Fermi-gas Feynman emulator, Nature Physics 8, 366 (2012)


\bibitem{ku}
Mark J. H. Ku, Ariel T. Sommer, Lawrence W. Cheuk, Martin W. Zwierlein,
Revealing the Superfluid Lambda Transition in the Universal Thermodynamics of a Unitary Fermi Gas,
 Science 335, 563-567 (2012).


\bibitem{mceoscompare}
J.E. Drut, T.A. Lahde, G. Wlazlowski, and P. Magierski, 
Equation of state of the unitary Fermi gas: an update on lattice calculations,
Phys. Rev. A {\bf 85}, 051601(R) (2012).


\bibitem{amherst}
N. V. ProkofÕev and B. V. Svistunov,
Bold diagrammatic Monte Carlo: A generic sign-problem tolerant technique for polaron models and possibly interacting many-body problems
Phys. Rev. B 77, 125101 (2008);
Nikolay ProkofÕev and Boris Svistunov, 
Fermi-polaron problem: Diagrammatic Monte Carlo method for divergent sign-alternating series,
Phys. Rev. B 77, 020408(R) (2008);
Nikolay ProkofÕev and Boris Svistunov,
Bold Diagrammatic Monte Carlo Technique: When the Sign Problem Is Welcome,
Phys. Rev. Lett. 99, 250201 (2007);
Emanuel Gull, David R. Reichman, and Andrew J. Millis,
Bold-line diagrammatic Monte Carlo method: General formulation and application to expansion around the noncrossing approximation,
Phys. Rev. B 82, 075109 (2010);
E. Gull, D. R. Reichman, and A. J. Millis, Phys. Rev. B {\bf 84}, 085134 (2011);
K. Van Houcke, F. Werner, N. Prokof'ev, B. Svistunov,
Bold diagrammatic Monte Carlo for the resonant Fermi gas,
arXiv:1305.3901.


\bibitem{oldgoulko}
O. Goulko and M. Wingate,
Thermodynamics of balanced and slightly spin-imbalanced Fermi gases at unitarity,
 Phys. Rev. A {\bf 82}, 053621 (2010)

\bibitem{lmc}
Michael Endres, David Kaplan, Jong-Wan Lee, and Amy Nicholson,
Lattice Monte Carlo calculations for unitary fermions in a finite box,
Phys. Rev. A {\bf 87} 023615 (2013).

\bibitem{af}
Bogdan Mihaila, John F. Dawson, Fred Cooper, Chih-Chun Chien, and Eddy Timmermans,
Auxiliary field formalism for dilute fermionic atom gases with tunable interactions,
Phys. Rev. A {\bf 83} 053637 (2011).


\bibitem{pollet}
Lode Pollet, Recent developments in Quantum Monte-Carlo simulations with applications for cold gases,
Rep. Prog. Phys. {\bf 75}, 094501 (2012).





\bibitem{contactmc}
. Van Houcke, F. Werner, E. Kozik, N. Prokof'ev, B. Svistunov, Contact and Momentum Distribution of the Unitary Fermi Gas by Bold Diagrammatic Monte Carlo,  arXiv:1303.6245.




\bibitem{realtime}
Guy Cohen, David R. Reichman, Andrew J. Millis, and Emanuel Gull,
GreenÕs functions from real-time bold-line Monte Carlo,
Phys. Rev. B {\bf 89},
115139 (2014).



\bibitem{magierski}
 P. Magierski, G. Wlazlowski, A. Bulgac, and J. E. Drut, 
 Finite-Temperature Pairing Gap of a Unitary Fermi Gas by Quantum Monte Carlo Calculations
 Phys. Rev. Lett. {\bf 103}, 210403 (2009); G. Wlazlowski, P. Magierski, J. E. Drut, A. Bulgac, and K. J. Roche,  
 Cooper Pairing Above the Critical Temperature in a Unitary Fermi Gas,
 Phys. Rev. Lett. {\bf 110}, 090401 (2013).


\bibitem{sheehy}
S. Q. Su, D. E. Sheehy, J. Moreno, and M. Jarrell, 
Dynamical cluster quantum Monte Carlo study of the single-particle
spectra of strongly interacting fermion gases
Phys. Rev. A {\bf 81}, 051604 (2010). 


\bibitem{virialsecond}
T.-L. Ho and E. J. Mueller, Phys. Rev. Lett. 92, 160404 (2004).



\bibitem{virialthird}
X.-J. Liu, H. Hu, P.D. Drummond, Phys. Rev. Lett. 102 (2009) 160401;



\bibitem{hightemp}
%X.-J. Liu, H. Hu, and P. D. Drummond, Phys. Rev. Lett. 102,
%160401 (2009);
K. M. Daily and D. Blume, 
Thermodynamics of the two-component Fermi gas with unequal masses at unitarity,
Phys. Rev. A {\bf 85}, 013609 (2012);
X. Leyronas, 
Virial expansion with Feynman diagrams
Phys. Rev. A {\bf 84}, 053633 (2011).
%;
%Vudtiwat Ngampruetikorn, Meera M. Parish, and Jesper Levinsen,
%High-temperature limit of the resonant Fermi gas,
%Phys. Rev. A 91, 013606 (2015)
%


\bibitem{virialfourth}
D. Rakshit, K.M. Daily, D. Blume, Phys. Rev. A 85 (2012) 033634;
S. Endo and Y. Castin, Phys. Rev. A 92, 053624 (2015);
Yangqian Yan and D. Blume, Path-Integral Monte Carlo Determination of the Fourth-Order Virial Coefficient for a Unitary Two-Component Fermi Gas with Zero-Range Interactions,
Phys. Rev. Lett. 116, 230401 (2016).


\bibitem{ngampruetikorn}
Vudtiwat Ngampruetikorn, Meera M. Parish, and Jesper Levinsen,
High-temperature limit of the resonant Fermi gas,
Phys. Rev. A {\bf 91}, 013606 (2015)



\bibitem{virialreview}
Xia-Ji Liu,
Virial expansion for a strongly correlated Fermi system and its application to ultracold atomic Fermi gases,
Physics Reports, {\bf 524}, 37 (2013)


\bibitem{virialcontact}
 H. Hu, X.-J. Liu, P.D. Drummond, 
Universal contact of strongly interacting fermions at finite temperatures,
New J. Phys. {\bf 13},  035007 (2011).


\bibitem{sunleyronas}
Mingyuan Sun and Xavier Leyronas,
High-temperature expansion for interacting fermions,
Phys. Rev. A {\bf 92}, 053611 (2015)

\bibitem{virialstructure}
Hui Hu and Xia-Ji Liu, Universal dynamic structure factor of a strongly correlated Fermi gas, 
Phys. Rev. A 85, 023612 (2012)


\bibitem{shenresponse}
Gang Shen, Response function of a strongly interacting Fermi gas in a virial expansion, Phys. Rev. A 87, 033612 (2013)







%\bibitem{bcsbecearly1}
%C. A. R. S‡ de Melo, M. Randeria, and J. R. Engelbrecht, Phys. Rev. Lett. {\bf 71}, 3202 (1993).
%\bibitem{bcsbecearly2}
%A. J. Leggett in {\em Modern Trends in the Theory of Condensed Matter}, ed. A. Pekalski and J. Przystawa (Springer, Berlin, 1980) p 14;
%D. M. Eagles, Phys. Rev. {\bf 186}, 456 (1969).
%\bibitem{nozieres}
%P. Nozieres and S. Schmitt-Rink, J. Low Temp. Phys, {\bf 59}, 195 (1985).
%







%
%\bibitem{otherpseudo}
%http://prb.aps.org/abstract/PRB/v83/i10/e104517




%%
%\bibitem{bruunreview}
%Massignan P, Zaccanti M and Bruun G M 2014 Rep. Prog. Phys. 77 034401


%
%\bibitem{janko}
%Boldizs\'ar Jank\'o, Jiri Maly, and K. Levin, Pseudogap effects induced by resonant pair scattering, Phys. Rev. B {\bf 56}, R11407 (1997).  Contrasts T-matrix models with phase fluctuation and preformed pairs.

%\bibitem{attractivehubbard}
%Mohit Randeria, Nandini Trivedi, Adriana Moreo, and Richard T. Scalettar,
%Phys. Rev. Lett. {\bf 69}, 2001 (1992);
%Nandini Trivedi and Mohit Randeria, Phys. Rev. Lett. {\bf 75}, 312 (1995).  Monte-Carlo study of attractive U Hubbard model.
%
%\bibitem{nguyen}Nguyen, H. Q. et al. Observation of giant positive magnetoresistance in a Cooper pair insulator. Phys. Rev. Lett. 103, 157001 (2009).
%\bibitem{sacepe}
%SacŽpŽ, B. et al. Disorder-induced inhomogeneities of the superconducting state close to the superconductorÐinsulator transition. Phys. Rev. Lett. 101, 157006 (2008).
%%
%\bibitem{barber}
%R. P. Barber, L. M. Merchant, A. LaPorta, and R. C. Dynes,
%Phys. Rev. B 49, 3409 (1994).









\bibitem{wilsonkogut}
K. G. Wilson and J. Kogut, Phys. Rep. 12, 75 (1974)

\bibitem{epsilon}
Y. Nishida and D. T. Son. $\epsilon$ expansion for a Fermi gas at infinite scattering length. Phys. Rev. Lett. {\bf 97}, 050403, (2006)

\bibitem{epsilon2}
Y. Nishida and D. T. Son. Fermi gas near unitarity around four and two spatial dimensions. Phys. Rev. A, {\bf 75}, 063617,  (2007).

\bibitem{nussinov}
Zohar Nussinov and Shmuel Nussinov,
Triviality of the BCS-BEC crossover in extended dimensions: Implications for the ground state energy,
Phys. Rev. A {\bf 74}, 053622 (2006).

\bibitem{nishida}
Yusuke Nishida, 
Unitary Fermi gas at finite temperature in the $\epsilon$ expansion,
Phys. Rev. A {\bf 75}, 063618 (2007).

\bibitem{cheneps}
Jiunn-Wei Chen and Eiji Nakano,
BEC-BCS crossover in the $\epsilon$ expansion,
Phys. Rev. A {\bf 75}, 043620 (2007).

\bibitem{moreeps}
Andrei Kryjevski, 
Effective Lagrangian of unitary Fermi gas from $\epsilon$ expansion,
Phys. Rev. A 78 043610 (2008).

\bibitem{arnold}
Peter Arnold, Joaqu\'in E. Drut, and Dam Thanh Son,
Next-to-next-to-leading-order $\epsilon$ expansion for a Fermi gas at infinite scattering length,
Phys. Rev. A {\bf 75}, 043605 (2007).


\bibitem{largeN}
M. Veillette, D. Sheehy, and L. Radzihovsky. Large-N expansion for unitary superfluid Fermi gases. Phys. Rev. A, {\bf 75}, 043614 (2007);
Hiroaki Abuki and Tomas Brauner,
Strongly interacting Fermi systems in 1/N expansion: From cold atoms to color superconductivity,
Phys. Rev. D {\bf 78}, 125010 (2008).

\bibitem{sachdevlargeN}
P. Nikolic and S. Sachdev. Renormalization-group fixed points, universal phase diagram, and 1/N expansion for quantum liquids with interactions near the unitarity limit. Phys. Rev. A, {\bf 75}, 033608 (2007).


\bibitem{earlyhaussmann}
R. Haussmann. Properties of a Fermi liquid at the superfluid transition in the crossover region between BCS superconductivity and Bose-Einstein condensation. Phys. Rev. B, {\bf 49}, 12975 (1994).

\bibitem{laterhaussmann}
R. Haussmann, W. Rantner, S. Cerrito, and W. Zwerger. Thermodynamics of the BCS-BEC crossover. Phys. Rev. A, {\bf 75}, 023610 (2007).

\bibitem{km}
Leo P. Kadanoff and Paul C. Martin,
Theory of Many-Particle Systems. II. Superconductivity,
Phys. Rev. {\bf 124}, 670 (1962).

\bibitem{bcsbecearly1}
C. A. R. S‡ de Melo, M. Randeria, and J. R. Engelbrecht, Phys. Rev. Lett. {\bf 71}, 3202 (1993).


\bibitem{chen2}
Chen, Q. J., I. Kosztin, B. Janko, and K. Levin, 1998, Phys. Rev. Lett. 81(21), 4708.
Chen, Q. J., I. Kosztin, B. Janko, and K. Levin, 1999, Phys. Rev. B 59(10), 7083.
Chen, Q. J., I. Kosztin, and K. Levin, 2000, Phys. Rev. Lett. 85(14), 2801.
Chen, Q. J., K. Levin, and I. Kosztin, 2001, Phys. Rev. B 63, 184519. Chen, Q. J., and J. R. Schrieffer, 2002, Phys. Rev. B 66, 014512.

%


 \bibitem{bcsbecpseudo}
C.-C. Chien, H. Guo, Y. He, and K. Levin, arXiv:0910.3699;
Q. C. Chen, and K. Levin, Phys. Rev. Lett. {\bf 102}, 190402 (2009).;
Q. Chen, J. Stajic, S. Tan, and K. Levin, Physics Reports {\bf 412}, 1 (2005);
G. Oritz and J. Dukelsky, Phys. Rev. A {\bf 72}, 403611 (2005);
A. Perali, P. Pieri, L. Pisani, and G. C. Strinati, Phys. Rev. Lett. {\bf 92}, 220404 (2004).

\bibitem{haussmann}
Haussmann, Punk, and Zwerger, Spectral functions and rf response of ultracold fermionic atoms,
Phys. Rev. A {\bf 80}, 063612 (2009). 
% Fully self-consistent diagrammatic model.
%arXiv:0904.1333 (2009).


\bibitem{kinnunenhartree}
J. J. Kinnunen, Hartree shift in unitary Fermi gases,
Phys. Rev. A 85, 012701 (2012);


 
 \bibitem{strinati}
P. Pieri, A. Perali, G. C. Strinati, Enhanced paraconductivity-like fluctuations in the radiofrequency spectra of ultracold Fermi atoms, Nature Physics {\bf 5}, 735 (2009).  
%Paraconductivity in the BCS-BEC crossover.


\bibitem{perali} A. Perali, P. Pieri, G. C. Strinati, and C. Castellani, Pseudogap and spectral function from superconducting fluctuations to the bosonic limit, Phys. Rev. B {\bf 66}, 024510 (2002). 

\bibitem{hu3}
H. Hu, X.-J. Liu, and P. D. Drummond, Europhys. Lett. 74, 574
(2006).

\bibitem{luhu}
Xia-Ji Liu and Hui Hu, Self-consistent theory of atomic Fermi gases with a Feshbach resonance at the superfluid transition,
Phys. Rev. A {\bf 72}, 063613 (2005);
X.-J. Liu and H. Hu, 
Equation of state of a superfluid Fermi gas in the BCS-BEC crossover,
Europhys. Lett. {\bf 75}, 364 (2006).

% \bibitem{hu}
%Hui Hu, Xia-Ji Liu, Peter D. Drummond, and Hui Dong, Phys. Rev. Lett. {\bf 104}, 240407 (2010). Quantum Cluster Expansion

\bibitem{hu2}
H. Hu, P. D. Drummond, and X.-J. Liu. Universal thermodynamics of strongly interacting Fermi gases. Nat. Phys. {\bf 3}, 469 (2007).
















%Two-Dimensional Fermi Liquid with Attractive Interactions
%B. Fršhlich, M. Feld, E. Vogt, M. Koschorreck, M. Kšhl, C. Berthod, and T. Giamarchi
%Phys. Rev. Lett. 109 130403 (2012)


%\bibitem{krinner}
%S. Krinner et al., PNAS 201601812 (2016)
%\bibitem{sheehyrf}
%Martin Veillette, Eun Gook Moon, Austen Lamacraft, Leo Radzihovsky, Subir Sachdev, and D. E. Sheehy,
%Radio-frequency spectroscopy of a strongly imbalanced Feshbach-resonant Fermi gas,
%Phys. Rev. A 78, 033614 (2008)

%\bibitem{drechsler}
%M. Drechsler and W. Zwerger, Crossover from BCS- superconductivity to Bose-condensation,
% Ann. Phys. {\bf 1}, 15 (1992)

\bibitem{howleclair}
Pye-Ton How and Andr\'e LeClair,
S-matrix approach to quantum gases in the unitary limit: II. The three-dimensional case,
J. Stat. Mech. {\bf 7}, 07001 (2010)

\bibitem{weiler}
Erik M. Weiler and Theja N. De Silva,
Thermodynamic properties of universal Fermi gases,
Phys. Rev. A {\bf 87}, 013602 (2013).

\bibitem{rg}
S. Diehl, H. Gies, J. Pawlowski, and C. Wetterich,
Renormalization flow and universality for ultracold fermionic atoms,
Phys. Rev. A {\bf 76}, 053627 (2007);
S. Diehl, H. Gies, J. Pawlowski, and C. Wetterich,
Flow equations for the BCS-BEC crossover,
Phys. Rev. A {\bf 76}, 021602(R) (2007).

\bibitem{morerg}
Igor Boettcher, Jan M. Pawlowski, and Christof Wetterich,
Critical temperature and superfluid gap of the unitary Fermi gas from functional renormalization,
Phys. Rev. A {\bf 89}, 053630 (2014).

\bibitem{brg}
Igor Boettcher, Jan M. Pawlowski, and Sebastian Diehl,
Ultracold atoms and the Functional Renormalization Group,
Nuc. Phys. B - Proc. Sup. {\bf 228}, 63 (2012).

\bibitem{gubbelsrg}
K. B. Gubbels and  H. T. C. Stoof,
Renormalization Group Theory for the Imbalanced Fermi Gas,
Phys. Rev. Lett. {\bf 100}, 140407 (2008).

\bibitem{bg}
Elmer V. H. Doggen and Jami J. Kinnunen,
Momentum-resolved spectroscopy of a Fermi liquid,
Scientific Reports, {\bf 5}, 9539 (2015).


\bibitem{ds}
S. Diehl and C. Wetterich, 
Universality in phase transitions for ultracold fermionic atoms,
Phys. Rev. A {\bf 73}, 033615  (2006);
S. Diehl and C. Wetterich, 
Functional integral for ultracold fermionic atoms,
Nucl. Phys. B {\bf 770}, 206  (2007);. 
Diehl, 
Functional renormalization for quantum phase transitions with nonrelativistic bosons,
Phys. Rev. B {\bf 77}, 064504 (2008).

\bibitem{bs}
A. Vagov, H. Schomerus, and A. Shanenko,
Generalized Galitskii approach for the vertex function of a Fermi gas with resonant interaction
Phys. Rev. B {\bf 76}, 214513 (2007).

\bibitem{cazalilla}
M. A. Cazalilla,
A Composite Fermion Approach To The Ultracold Dilute Fermi Gas,
Int. J. Mod. Phys. B {\bf 25}, 329 (2011).


\bibitem{opepseudo}
Philipp Gublera,  Naoki Yamamotoc, Tetsuo Hatsudab, and Yusuke Nishida,
Single-particle spectral density of the unitary Fermi gas: Novel approach based on the operator product expansion, sum rules and the maximum entropy method,
Annals of Physics, {\bf 356}, 467 (2015).

\bibitem{effective}
Boris Krippa, Effective field theory and cold Fermi gases near unitary limit, Phys. Rev. A {\bf 76}, 053622 (2007);
Boris Krippa, Exact renormalization group flow for ultracold Fermi gases in the unitary limit,
J. Phys. A: Math. Theor. {\bf 42}, 465002 (2009).

\bibitem{pieristrinati}
P. Pieri and G. C. Strinati,
Strong-coupling limit in the evolution from BCS superconductivity to Bose-Einstein condensation,
Phys. Rev. B {\bf 61}, 15370 (2000).

\bibitem{pierit}
P. Pieri, L. Pisani, and G. C. Strinati, 
BCS-BEC crossover at finite temperature in the broken-symmetry phase,
Phys. Rev. B {\bf 70}, 094508  (2004).

\bibitem{micnast}
R. Micnas, 
On the Crossover from BCS Superconductivity to Bose Condensation 
Acta Phys. Pol. A {\bf 100s}, 177  (2001). 

\bibitem{gubelstoof}
K. B. Gubbels and H. T. C. Stoof,
Interacting preformed Cooper pairs in resonant Fermi gases,
Phys. Rev. A {\bf 84}, 013610 (2011),

\bibitem{mulkerin}
Brendan C. Mulkerin, Xia-Ji Liu, and Hui Hu,
Beyond Gaussian pair fluctuation theory for strongly interacting Fermi gases,
Phys. Rev. A {\bf 94}, 013610 (2016).
%
%\bibitem{twod}
%
%\bibitem{imb}
%K. B. Gubbels, and H. T. C. Stoof, 
%Imbalanced Fermi gases at unitarity,
%Physics Reports, {\bf 525}, 255 (2013)

\bibitem{mottmetal}
M. Imada, A. Fujimori, and Y. Tokura, Metal-insulator transitions, Rev. Mod. Phys. {\bf 70}, 1039 (1998) .

\bibitem{oned}
Yean-an Liao, Ann Sophie C. Rittner, Tobias Paprotta, Wenhui Li, Guthrie B. Partridge, Randall G. Hulet, Stefan K. Baur, Erich J. Mueller, Spin-Imbalance in a One-Dimensional Fermi Gas, Nature {\bf 467}, 567 (2010)

\bibitem{fgm}
Elmar Haller,	James Hudson,	Andrew Kelly,	Dylan A. Cotta,	Bruno Peaudecerf,	Graham D. Bruce, and
 Stefan Kuhr,
Single-atom imaging of fermions in a quantum-gas microscope
Nature Physics {\bf 11}, 738 (2015);
Lawrence W. Cheuk, Matthew A. Nichols, Melih Okan, Thomas Gersdorf, Vinay V. Ramasesh, Waseem S. Bakr, Thomas Lompe, and Martin W. Zwierlein,
Quantum-Gas Microscope for Fermionic Atoms,
Phys. Rev. Lett. {\bf 114}, 193001 (2015);
Maxwell F. Parsons, Florian Huber, Anton Mazurenko, Christie S. Chiu, Widagdo Setiawan, Katherine Wooley-Brown, Sebastian Blatt, and Markus Greiner,
Site-Resolved Imaging of Fermionic $^6$Li
 in an Optical Lattice
Phys. Rev. Lett. {\bf 114}, 213002 (2015);
G. J. A. Edge, R. Anderson, D. Jervis, D. C. McKay, R. Day, S. Trotzky, J. H. Thywissen,
Imaging and addressing of individual fermionic atoms in an optical lattice,
Phys. Rev. A {\bf 92}, 063406 (2015);
A. Omran, M. Boll, T. A. Hilker, K. Kleinlein, G. Salomon,
I. Bloch, and C. Gross, 
Microscopic Observation of Pauli Blocking in Degenerate Fermionic Lattice Gases,
Phys. Rev. Lett. {\bf 115}, 263001
(2015).


\bibitem{pwave}
Y. Ohashi,
BCS-BEC Crossover in a Gas of Fermi Atoms with a p-Wave Feshbach Resonance,
Phys. Rev. Lett. {\bf 94}, 050403  (2005);
M. Iskin and C. A. R. S‡ de Melo,
Evolution from BCS to BEC Superfluidity in p-Wave Fermi Gases,
Phys. Rev. Lett. {\bf 96}, 040402 (2006);
Daisuke Inotani, Ryota Watanabe, Manfred Sigrist, and Yoji Ohashi,
Pseudogap phenomenon in an ultracold Fermi gas with a p-wave pairing interaction,
Phys. Rev. A {\bf 85}, 053628 (2012);
J. P. Gaebler, J. T. Stewart, J. L. Bohn, and D. S. Jin,
p-Wave Feshbach Molecules,
Phys. Rev. Lett. {\bf 98}, 200403 (2007)

\bibitem{higherspin}
T.B. Ottenstein, T. Lompe, M. Kohen, A.N. Wenz, and S. Jochim, 
Collisional Stability of a Three-Component Degenerate Fermi Gas,
Phys. Rev. Lett. {\bf 101},
203202 (2008);
J.H. Huckans, J.R. Williams, E.L. Hazlett, R.W. Stites, and K. M. OÕHara, 
Three-Body Recombination in a Three-State Fermi Gas with Widely Tunable Interactions
Phys. Rev. Lett.
{\bf 102}, 165302 (2009).

\bibitem{longrange}
M.A. Baranov, Theoretical progress in many-body physics with ultracold dipolar gases,
Phys. Rep. {\bf 464}, 71 (2008);
M. A. Baranov, M. Dalmonte, G. Pupillo, and P. Zoller,
Condensed Matter Theory of Dipolar Quantum Gases,
Chem. Rev. {\bf 112}, 5012 (2012).



\end{thebibliography}
\end{document}